\documentclass[aps,prd,groupedaddress,superscriptaddress]{revtex4-2}
\usepackage[utf8]{inputenc}
\usepackage{lineno}
\usepackage{rotating}
\title{Rare And Precision Frontier Report}
\date{\today}
\usepackage{url}
\usepackage[]{fancyhdr} 
\input{common/preamble}
\usepackage{relsize}
\def\babar{\mbox{\slshape B\kern-0.1em{\smaller A}\kern-0.1em
    B\kern-0.1em{\smaller A\kern-0.2em R}}}
\newcommand\snowmass{\begin{center}\rule[-0.2in]{\hsize}{0.01in}\\\rule{\hsize}{0.01in}\\
\vskip 0.1in Submitted to the  Proceedings of the US Community Study\\ 
on the Future of Particle Physics (Snowmass 2021)\\ 
\rule{\hsize}{0.01in}\\\rule[+0.2in]{\hsize}{0.01in} \end{center}}

\usepackage{subfiles}
\usepackage{booktabs}
\usepackage{xcolor}
\usepackage{hyperref}
\hypersetup{
    colorlinks=true,
    linkcolor=blue,
    filecolor=magenta,      
    urlcolor=blue
    }
\begin{document}
\title{Report of the Frontier for Rare Processes and Precision Measurements}
\date{\today}
\author{ \textbf{Frontier Conveners:} Marina~Artuso}\affiliation{Physics, Syracuse University, Syracuse, NY, 13244, USA}
\author{Robert~H~Bernstein}\affiliation{Muon Department, Fermilab, Batavia, IL, 60510, USA}
\author{Alexey~A~Petrov\vspace{0.1in}}\affiliation{Department of Physics and Astronomy, University of South Carolina, Columbia, SC, 29208, USA}
\affiliation{Department of Physics and Astronomy, Wayne State University, Detroit, MI, 48154, USA}
\author{\\ \textbf{Topical Group Conveners:} Thomas~Blum}\affiliation{Department of Physics, University of Connecticut, Storrs, CT, 06269-3046, USA}
\author{Angelo~Di Canto}\affiliation{Physics Department, Brookhaven National Laboratory, Upton, NY, 11973, USA}
\author{Sacha~Davidson}\affiliation{Department of Physics,	IN2P3 CNRS, Montpellier, 34095, France}
\author{Bertrand~Echenard}\affiliation{Department of Physics, Mathematics and Astronomy, California Institute of Technology,	Pasadena, CA, 91125, USA}
\author{Stefania Gori}\affiliation{Department of Physics,	University of California, Santa Cruz, Santa Cruz, CA, 95064, USA}
\author{Evgueni~Goudzovski}\affiliation{School of Physics and Astronomy, University of Birmingham, Birmingham,  B15 2TT, United Kingdom}
\author{Richard~F~Lebed}\affiliation{Department of Physics, Arizona State University, Tempe, AZ, 85287-1504, USA}
\author{Stefan Meinel}\affiliation{Department of Physics, University of Arizona, Tucson, AZ, 85721, USA}
\author{Emilie~Passemar}\affiliation{Department of Physics and Center for Exploration of Energy and Matter, Indiana University, Bloomington, IN, 47401, USA}\affiliation{Theory Center, Thomas Jefferson National Accelerator Facility, Newport News, VA, 23606, USA}
\affiliation{Física Te\`{o}rica	IFIC, Universitat de Val\`{e}ncia-CSIC, Valencia, 46071, Spain}
\author{Pavel~Fileviez Perez}\affiliation{Department of Physics, Case Western Reserve University, Cleveland, OH, 44106, USA}
\author{Tomasz~Skwarnicki}\affiliation{Physics, Syracuse University, Syracuse, NY, 13244, USA}
\author{Andrea Pocar}\affiliation{Department of Physics, University of Massachusetts, Amherst, Amherst, MA, 01003, USA}
\author{Mike~Williams}\affiliation{Laboratory for Nuclear Science, MIT, Cambridge, MA, 02139, USA}
\author{Peter~Winter\vspace{0.1in}}\affiliation{High Energy Physics Division, Argonne National Laboratory, Lemont, IL, 60439, USA}
\author{\\ \textbf{Liaisons:} 
Joshua~L~Barrow}
\affiliation{Department of Physics, MIT, Cambridge, MA, 02139, USA}
\affiliation{Department of Physics, Tel Aviv University, Tel Aviv, Israel}
\affiliation{Department of Physics, University 
of Tennessee, Knoxville, Tennessee, 37996, USA}
\author{Jake~Bennett}\affiliation{Physics and Astronomy, University of Mississippi, University, MS, 38677, USA}
\author{Susan~Gardner}\affiliation{Department 
of Physics and Astronomy, University of Kentucky, Lexington, 
Kentucky, 40506-0055, USA}
\author{Sophie~C~Middleton}
\affiliation{Department of Physics, Mathematics and Astronomy, California Institute of Technology,	Pasadena, CA, 91125, USA}
\author{Eric~Prebys}\affiliation{Department of Physics, University of California, Davis, Davis, CA, 90095, USA}
\author{Manuel Franco Sevilla}\affiliation{Department of Physics, University of Maryland, College Park, MD, 20742, USA}
\maketitle
\snowmass{}\pagebreak
\tableofcontents

\pagestyle{fancy}\lhead{}
\section{Executive Summary}

\subsection{Motivation and Big Questions}

The Rare Processes and Precision Measurements Frontier, referred to as the ``Rare and Precision Frontier", or RPF, encompasses searches for extremely rare processes or tiny deviations from the Standard Model (SM) that can be studied with intense sources and high precision detectors.  Our community studies have identified several unique research opportunities that may pin down the scales associated with New Physics (NP) interactions  and constrain the couplings of possible new degrees of freedom.  Studies of rare flavor transition transitions and precision measurements are indispensable  probes of flavor and fundamental symmetries, and provide insights into physics that manifests itself at higher energy or through weaker interactions than those directly accessible at high-energy colliders.

The Frontier explores the following fundamental topics:
\begin{itemize}
\item
the origin of quark and lepton flavor, generations, and mass hierarchies;
\item
the exploitation of flavor (both quark and lepton) as a precision probe of the Standard Model; 
\item 
the use of flavor physics as a tool for discovering new physics;
\item
the origin of the fundamental symmetries and their breakdown mechanisms;
\item
the physics of the dark sector available at high-intensity machines;

\item
the origins of baryon and lepton number violation, through the investigation of processes such 0$\nu\beta\beta$ decays, proton decays, or baryon-antibaryon oscillations
\item searches for non-zero electric dipole moments (EDMs) and CP-violation as well as  fundamental (for example,  Lorentz) symmetry tests;
\item
the ways in which non-perturbative Quantum Chromodynamics (QCD) explains the rich landscape of  hadron spectroscopy, including both conventional and exotic hadrons.
\end{itemize}
The opportunities presented in this report form an open ecosystem with intertwined research goals. These goals are  motivated by the Particle Physics Projects Prioritization Panel (P5) driver ``explore the unknown,'' which  connects all of the physics frontiers. Studies of neutrinos, quarks, and charged leptons provide complementary information, which can lead to a more complete picture of the physics underlying the question of the generations and the different properties associated with the flavor quantum numbers. Why are there three generations of quarks, charged leptons, and neutrinos, and what links the flavor of quarks to the flavor of leptons? Both quarks and leptons have flavor and mix, but the matrices describing the couplings between different flavors, the Cabibbo-Kobayashi-Maskawa (CKM) matrix for quarks and the Pontecorvo-Maki-Nakagawa-Sakata (PMNS) for neutrinos, have  different structures. The CKM matrix is highly hierarchical, with the diagonal elements with magnitude close to unity, and progressively smaller off-diagonal elements, as seen in the Wolfenstein parameterization~\cite{PhysRevLett.51.1945}.  In contrast,  the PMNS matrix has much larger off-diagonal elements.  Charged leptons do not mix like quarks in the Standard Model. These patterns of mixing for the charged and neutral fermions are a mystery.  The P5 driver ``Pursue the physics associated with neutrino mass" does not explicitly mention neutrino flavor or any potential linkage to the quark sector, leaving out a set of crucial questions in neutrino physics. Reaching outside the quarks and leptons, we can also ask, ``do leptonic decays of the Higgs violate flavor?"

 Flavor plays a critical dual role in high-energy physics.  First, the origin of the flavor structure of the Standard Model and various hierarchies of quark and lepton properties have been a fundamental unanswered question since the discovery of the muon. The limits on ${\mathcal B}(\mu \rightarrow e \gamma)$ led to the prediction that more than one neutrino species must exist.  Rare decays and flavor transitions offer many opportunities to probe the structure of possible new physics.  Second, flavor transitions are a powerful tool in constraining (and suggesting) possible models of Beyond the Standard Model (BSM) physics in light of the richness of the observed decay modes. Flavor is thus a common theme across much of the Frontier. Investigating flavor combines theoretical and experimental research pursued through a combination of small, medium, and large experiments that complement and enrich this broad program. Placing all of these closely related topics into a general ``explore the unknown" driver  captures neither how important they are nor how linked they are, both among themselves and across the Frontiers. In order to capture the unique opportunities provided by the study of physics that distinguishes different flavors, we proposed to introduce a new science driver, {\it flavor as a tool for discovery}.

Our Frontier also investigates the conservation of baryon and lepton numbers, captured in the P5 driver of ``new physical principles."  Proton decay and the $\Delta L = 2$ process of neutrinoless double beta decay are shared by the Neutrino and Rare and Precision Frontiers, but this Frontier also examines other baryon and lepton-number violating processes, such as $n\bar{n}$ oscillations.  

Another P5 driver was ``identify the new physics of dark matter." The Rare and Precision Frontier studies the dark sector using intense beams to discover dark matter (DM) and dark sector  particles, generally in the electron-to-proton mass range.  The interactions of dark matter particles created by thermal processes with  visible matter in the hot early universe could  explain its abundance today. We are poised to begin a simultaneously focused and wide-ranging period of exploration, with specialized experiments at colliders, searches in the flavor factories,  and dedicated experiments that will either discover such particles or improve the limits by factors of 10 to 1000.

\subsection{Community Consensus}

This report covers many invaluable physics topics, and we encourage the reader to explore them. We have identified seven central points for the broader community (unordered and intertwined):
\begin{itemize}
\item  We call for a new science driver, {\it flavor as a tool for discovery}, with two main goals. The first objective aims at uncovering the underlying reason for family replication and their different properties. The second encompasses the systematic study of flavor-specific decays that probe new physics through its interference with SM amplitudes. It is important to note that minimum flavor violation is not necessarily applicable to several plausible NP models. The effort  to understand the physics of flavors and generations is central to both physics programs articulated by the Neutrino and Rare Processes and Precision Measurements Frontiers, but this centrality needs to be articulated more clearly.  
\item The U.S. should support the LHCb Phase-II upgrades and Belle II. These experiments are broad, powerful, and irreplaceable probes of flavor physics along with topics relevant to all the Frontiers. These experiments pursue complementary research programs that utilize different beams and detection techniques to explore a vast array of new physics in beauty, charm, and rare tau decays. In addition, they investigate the patterns of bound states as the manifestation of the richness of QCD in its non-perturbative regime and contribute significantly to the exploration of the dark sector.
\item We should select a portfolio of accelerator-based dark sector experiments that are well-motivated, unique, and affordable.  Many possibilities have been identified and studied during the Snowmass process. P5 needs to support this physics, with a consequent process through which  the community, DOE, and NSF select an efficient and effective subset of the opportunities.
\item Experiments investigating charged lepton flavor violation  and lepton number violation in the muon sector probe mass scales far beyond the direct reach of colliders, as well as explore the nature of the flavor physics issues core to HEP.  PIP-II at FNAL enables a new muon program at an unprecedented intensity that could increase the discovery potential of such experiments by at least an order of magnitude. The U.S.\ should support a vigorous R\&D program towards its realization.
\item The theory efforts that guide and enable these investigations, while not a Project, should be vigorously supported by P5. Our Frontier  relies on the techniques of  and guidance from Effective Field Theories (EFTs). Calculations of quantities throughout the Rare and Precision Frontier rely on computational methods that provide precise and unbiased results that can be systematically improved as needed, such as Lattice QCD. Precision measurements of deviations from the Standard Model, instead of the direct observation of new particles, require these methods' intellectual framework and power. In addition, theory-experiment collaboration in developing new experimental approaches has been at the foundation of much of the progress in dark sector searches and continues to be vital to this growing field.

\item A portfolio of experiments of different cost and time scales is an integral part of our physics program. Such experiments include: 
\begin{itemize}
\item experiments measuring electric dipole moments, in particular the proton EDM measurement in a storage ring, along with experiments exploiting synergies with AMO techniques to examine fundamental symmetries;
    \item experiments probing rare light meson decays, such as JEF \cite{pro-JEF14,pro-JEF17} and REDTOP~\cite{REDTOP:2022slw};
    \item involvement of the US experimental community in the vibrant international program at both CERN and J-PARC studying rare $K$ decays;
    \item the PIONEER experiment at  PSI will study lepton universality in the muon and electron generations, and is a  promising medium-scale experiment that should be supported.
\end{itemize}

\item We also stress that small and medium-sized multipurpose experiments allow  early career researchers to gain experience in many  stages of an experiment, from proposal writing to design and construction to publication. Examples are the Fermilab $g {-} 2$ or the quark flavor experiments, which allow junior scientists to participate in  instrument construction or computationally intensive tasks, as well as develop significant  physics analysis expertise.  In addition, the relatively small sizes of these experiments allow for less hierarchical organization, with a broad array of leadership opportunities. A commitment to mentor  early career scientists in a supportive and inclusive environment is guiding our programmatic choices.
\end{itemize}

\subsection{Physics of the Frontier and the P5 Drivers \label{sec:physicsDrivers}}

The Rare and Precision Frontier covers a broad range of topics, defined by neither a particle nor a machine. It is broadly aligned with the Intensity Frontier in the 2013 Community Summer Study~\cite{Hewett:2014qja}, except for the removal of the physics of neutrinos, which now is the focus of a separate Frontier. During the last P5 cycle, many of the various physics topics encompassed by our Frontier fell into the general  ``Explore the unknown: new particles, interactions, and physics principles" driver~\cite{HEPAPSubcommittee:2014bsm}: this broad motivation, which is the strong connecting theme of all of the efforts current and planned in the particle physics community, is correctly identified as taking two basic forms:  producing new particles and detecting
the quantum influence of new particles. Many experimental efforts discussed below fall into the latter category. Before we summarize how the RPF program  contributes to the physics drivers identified in previous community studies, we will articulate the need for a new physics driver centered around the physics of flavor. This driver puts  many theoretical and experimental advances over the last decade in proper focus and defines clearer objectives for the future.

\subsubsection{A new physics driver: flavor physics as a tool for discovery}

``Flavor physics" is defined as physics that distinguishes the generations.  We see three generations of quarks, and three generations of leptons. This replication, combined with the unexplained pattern of masses of quarks and leptons, which spans several orders of magnitude, is a central mystery of particle physics. Both quarks and leptons exhibit inter-generational mixing, with different patterns for the quarks (the CKM matrix) and the neutral leptons, or neutrinos (the PMNS matrix.) The reason for this difference is still unknown, but we must answer these questions in order to form a coherent picture of the elementary particles and their interactions. In addition, the richness of decays of different flavor species available to experimental observation allows us to probe deviations from SM expectations with a multitude of approaches that we will summarize below. Here the increasing precision of many ongoing and proposed experiments already shows some intriguing anomalies and may provide further clues to the nature and scale of the new physics. 

Much of our Frontier centers on these two themes.  Weak decays of all the quarks and the CKM matrix are subtle probes of deviations of Standard Model expectations and already constrain reasonably well the number of generations --- precision measurements of the CKM triangle and tests of unitarity are at the core of the Frontier. Rare decays seek to uncover new physics manifesting itself through interference with Standard Model diagrams. Searches for charged lepton flavor violation examine the nature of lepton flavor and, through $\Delta L = 2$ processes, extend our flavor studies to the sources of leptogenesis.  Whether in the $B$, $D$ or $K$ systems, comparisons of decay rates into different lepton species challenge the notion of lepton universality, strongly implied by the Standard Model. Hence flavor studies are either at the core or crucial tools across our Frontier, and the nature of flavor is a basic question in particle physics.

\subsubsection{Relation to 2014 P5 Drivers}
We next turn to the connections between the different physics topics of the Frontier and the P5 drivers identified after the 2013 community summer study. First, we aim at spelling out these relations. Second, we show how the Frontier is related to the other Frontiers. Third, we show how a full understanding of the puzzles motivating particle physics require the study of both flavor-dependent and flavor-independent phenomena. As a reminder, the 2014 P5 Drivers were:

\begin{itemize}

\item Use the Higgs boson as a new tool for discovery 
\item Pursue the physics associated with neutrino mass 
\item Identify the new physics of dark matter 
\item Understand cosmic acceleration: dark energy and inflation
\item Explore the unknown: new particles, interactions, and physical principles 
\end{itemize}

We have relatively little direct connection to cosmic acceleration and focus on the other drivers.

\subsubsection{Use the Higgs boson as a new tool for discovery}

The Rare and Precision Frontier's purview includes using precise measurements to search for deviations from the Standard Model. Intellectually, precision Higgs physics is part of the Rare and Precision Frontier. Functionally, such measurements have been assigned to the Energy Frontier because Higgs factories, such as the proposed FCC-ee, are colliders, and colliders are assigned to the Energy Frontier. 

Since much of physics objective of the Rare and Precision Frontier relies on indirect measurements, we provide examples of how RPF measurements depend on Higgs properties. In the interest of length, we only mention a few highlights and refer the reader to the Topical Group reports for a more in-depth discussion~\cite{https://doi.org/10.48550/arxiv.2208.05403,https://doi.org/10.48550/arxiv.2209.07156,https://doi.org/10.48550/arxiv.2209.00142}.

Lepton universality tests which could be related to the Higgs properties abound in $b$ and $c$ decays because of the dependence on the Higgs coupling on mass.   A charged Higgs would couple much more strongly to the $\tau$ than the muon or electron. Ratios of $b \rightarrow c \tau \bar{\nu}_{\tau}$ to $b \rightarrow c \ell^- \bar{\nu}_{l}$, usually called $R(X_c)$, where $X_c$ is the charmed hadron in the final state, are a natural place to search. The world averages of results for $R(D)$ and $R(D^*)$ exceed the Standard Model prediction with a combined significance of $\sim 3\sigma$, pointing to possible violations of lepton flavor universality~\cite{https://doi.org/10.48550/arxiv.2208.05403}.  A composite Higgs could be responsible for the violation of flavor universality possibly seen in $B^+ \rightarrow K^+ \ell^+\ell^-$~\cite{Niehoff_2015}.

Lepton-flavor violating Higgs decays have been the subject of considerable attention~\cite{https://doi.org/10.48550/arxiv.2209.00142}.  Searches for LFV decays of the Higgs  have been performed by the ATLAS, CMS, and LHCb experiments. In addition, searches for  Higgs decays into $e\tau$ or $\mu \tau$ pairs have been performed for different $\tau$ decays channels (and the FCC-ee could perform similar searches for $Z$ decays.)  Triplet Higgs models can predict lepton flavor violation, especially in $\mu \rightarrow 3e$~\cite{Kakizaki_2003}.

\subsubsection{Pursue the physics associated with neutrino mass}

The RPF topical group focused on baryon and lepton number violation shares an interest on $0\nu2\beta$ decays  with the Neutrino Frontier.
The question of whether neutrino is its own antiparticle has important implications for the matter-antimatter asymmetry of the Universe, as many scenarios of leptogenesis require neutrinos to be Majorana particles.

The physics associated with neutrino mass should certainly include the understanding of the PMNS matrix. The driver does not explicitly mention the structure of this matrix, but the reasons for neutrino flavor mixing and the non-diagonal nature of the matrix, in contrast to the relatively diagonal CKM matrix, indicate the physics of neutrino flavor differs in unexplained ways from the physics of quark flavor. In addition, some scenarios of leptogenesis require CP-phases of the PMNS matrix for the explanation of the observed matter-antimatter asymmetry. 

\subsubsection{Identify the new physics of dark matter}

The RPF has a vibrant community focused on a new set of intensity-frontier experiments that will offer crucial insights into the physics of dark sectors,
highlighted as a \textit{Priority Research Direction} in the 2018 ``Dark Matter New Initiatives" study~\cite{osti1659757}. The types of dark matter examined in the Frontier are roughly in the range between the electron and proton mass with couplings to the SM consistent with what was needed for thermal production in the early Universe. Accelerator-based dark-sector experiments include multipurpose detectors at colliders and flavor factories, along with dedicated fixed-target experiments and downstream detectors~\cite{Gori:2022vri,rpf6}.

\subsubsection{Explore the unknown: new particles, interactions, and physical principles }

Every physics Frontier explores the unknown and searches for new phenomena.  Nevertheless, the RPF makes unique contributions.  For example, we have already discussed the physics of flavor as a unique focus of the RPF: studies of the CKM matrix, charged lepton flavor violation, and much of the physics of weak decays belong to the RPF. Some studies have synergies with other frontiers. Other inquiries, such as tests of fundamental symmetries in small experiments, are  unique to the RPF.  The RPF studies:
\begin{itemize}
\item electric dipole moments and new sources of CP violation ranging from fundamental particles to molecules and using both small-scale AMO methods as well as storage rings; 
\item rare decay modes of charm and beauty hadrons~\cite{https://doi.org/10.48550/arxiv.1012.3893}, as well as rare kaon decays, such as $K_L \rightarrow \pi \ell\ell $ or $K \rightarrow \pi \nu \bar{\nu}$, which are exceptionally stringent tests of the Standard Model;
\item magnetic dipole moments of the electron, muon and tau leptons;
\item flavor universality, in beauty and charm decays, in  $\tau$ decays, and in precision tests using $\pi$ and $K$ decays;
\item violations of Lorentz symmetries and precision tests of gravity.
\end{itemize}

These signature searches, combined with the physics of flavor, expand the RPF far beyond the ``Intensity Frontier." The breadth of tools for new physics searches

perhaps single out the RPF as the Frontier most directly related to the exploration of the unknown.

\section{Introduction}

Particle physics tries to understand the innermost structure of the Universe and the dynamics underlying its origin and evolution. Enormous strides have been made in the last few decades, with a solid body of work confirming the beauty and power of the theory that we call the Standard Model, which is able to describe and predict countless observations, in some cases with impressive accuracy. Yet we strive for a deeper knowledge, motivated by puzzles that still await a more complete explanation. The physics topics we grappled with in our Frontier are closely intertwined with some of these profound questions. For example, the stability of the Universe we live in is rooted on the asymmetry between matter and antimatter abundances. Sakharov identified three conditions necessary for this occurrence\cite{Sakharov:1967dj}: baryon number violation, new sources of C and CP violation, and interactions out of thermal equilibrium. Several of the studies that we consider have unique sensitivities to new sources of CP violation and baryon number violation. Another important example is the elusive nature of dark matter, which is more abundant than ordinary matter and whose nature remains unknown: its only well-measured property is its gravitational interaction with ordinary matter. The study of a possible dark sector interacting with ordinary matter via portals can shed light on some plausible dark matter candidates. 

\begin{figure}
    \centering
    \includegraphics[width=1.0\textwidth]{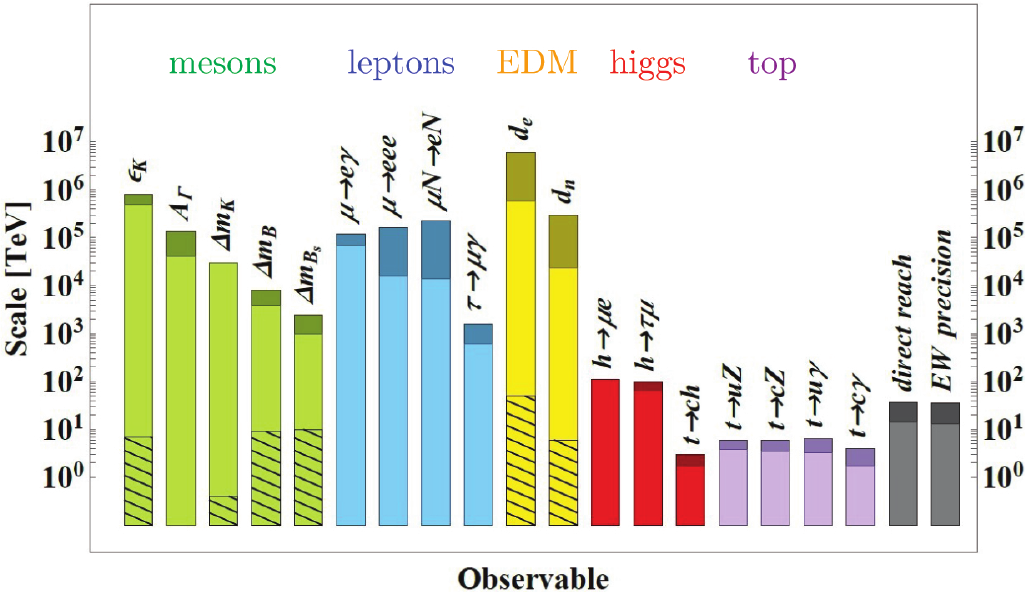}
    \caption{Reach in new physics of present and future facilities, from generic dimension-six operators. Different probes are identified by color coding: green is for mesons, blue for leptons, yellow for EDMs, red for Higgs flavored coupling, and purple for the top quark. The grey columns illustrate the reach of direct searches and electroweak precision studies. The coupling coefficients of these operators are taken to be of ${\cal O}(1)$ in the solid color columns or suppressed by MFV factors (hatch-ﬁlled surfaces). Light colors correspond to present data, and dark colors correspond to mid-term prospects in the time scale of the HL-LHC~\cite{EuropeanStrategyforParticlePhysicsPreparatoryGroup:2019qin, https://doi.org/10.48550/arxiv.2209.10639}.
        \label{fig:mass-scales}}
\end{figure}

While we have compelling reasons to expect that new particles and interactions are needed to achieve a more complete picture of the microworld, the mass scale of this new physics is not known. It is possible that it is light and very weakly interacting, or it is possible that it is much higher than the $\approx $ TeV scale originally anticipated. Precision measurements, probing quantum effects that allow heavy new particles to manifest themselves through their effects on rare and forbidden decays in the SM paradigm, exploit large data sets and precise instruments to probe mass ranges beyond the direct reach of high-energy machines. Fig.~\ref{fig:mass-scales} illustrates the mass scale reach of some of the measurements discussed in this report, and much useful discussion can be found in Ref.~\cite{https://doi.org/10.48550/arxiv.2209.10639}.

The report is organized as follows.  First, Section~\ref{sec:tg} describes how the RPF study was organized into seven areas considered by corresponding working groups.  Next, in Section~\ref{sec:exp-methods}, we  discuss the current experimental status of the broad suite of rare processes and precision measurements examined in our study, and the ways in which future improvements may be achieved in the medium- and long-term future, with upgrades to existing facilities and experiments or new initiatives. Section~\ref{sec:rd} summarized the R\&D efforts needed to support this experimental program.
Section~\ref{sec:theory} covers the theoretical landscape, and how different theoretical approaches can illuminate or guide experimental studies. Section~\ref{sec:connections}  elucidates the connections with other frontiers and experimental programs, and we review the impact of our work on education and society. Finally, we conclude with the big ideas emerging from this study.

\section{Major themes of the RPF topical groups}
\label{sec:tg}

The broad physics landscape examined by the RPF is organized into seven topical groups, who articulated its fundamental components. Here we summarize the main themes explored in each topical group.

\subsection{Weak decays of {\bf {\it b}} and {\bf {\it c}} quarks}

Studies of heavy-flavored hadrons provide rich, diverse, and model-independent probes for new physics at energy scales far beyond what is directly accessible. Heavy-flavor physics is crucial to our search for new physics in the upcoming 10--20 years, important advances are expected through a highly synergistic program of experiments: LHCb and its planned upgrades at the LHC, Belle II at the SuperKEKB asymmetric $e^+e^-$ collider, and the $e^+e^-$ charm factories BESIII and STCF. In particular, Belle II and LHCb have the unique potential to unveil new physics by confirming intriguing hints of deviations from the Standard Model that have been recently observed in $b \rightarrow s \ell^+ \ell^-$ and $b \rightarrow c \tau\bar{\nu}$  transitions at \babar, Belle, and LHCb, or finding new unexpected outcomes in the study of rare and forbidden decays such as $b\rightarrow s \nu \bar{\nu}$ or $B_{(s)}\rightarrow \ell^+\ell^-$. In addition, the continuing refinement of the studies of quark mixing, flavor oscillations and CP violation may unveil subtle deviations from SM expectations. Farther into the future, the experimental program can be extended at the $e^+e^-$ circular colliders  proposed for precision studies of the Higgs boson. The U.S.\  flavor community is well-positioned to lead key aspects of the physics, computing, and detector construction in all of these experimental programs. Theoretical efforts that inspire and elucidate the fundamental impact of experimental findings are crucial to success and should be supported.
The lepton universality and other stringent tests foreseen in these experiments and their upgrades are discussed in the $b$ and $c$ topical group report~\cite{https://doi.org/10.48550/arxiv.2208.05403}.

\subsection{ Weak decays of the light quarks}

Studies of the light quarks include precision measurements of both flavor-conserving and flavor-violating decays of kaons, hyperons, and $\eta/\eta^\prime$ mesons. Tests of new physics through checks of the unitarity of the first column  of the CKM matrix, along with lepton flavor/number and lepton universality tests, have revealed experimental anomalies. These findings require additional experimental and theoretical studies of the light quark systems to conclusively assess their impact. Synergy with the studies of the heavy quark systems is assured.

The study of $K$ physics is centered in Europe and Asia, but a small and vibrant community in the US is contributing to this important program. Such experiments are natural complements to searches of new physics in $b$-decays and provide constraints to the CKM unitarity triangle.
New ideas exploit rare pion decays, such as PIONEER, which plans to study the ratio $R_{e/\mu} = \Gamma(\pi^+ \rightarrow e^+ \nu_e)/ \Gamma (\pi^+ \rightarrow \mu^+ \nu_{\mu}) $ to 0.01\% precision, probing up to PeV mass scales\cite{https://doi.org/10.48550/arxiv.2203.01981}. In later stages, the experiment could study pion beta decay ($\pi^+ \rightarrow \pi^0 e^+ (\gamma)$), testing CKM universality through the precision measurement of $|V_{ud}|$. Finally, $\eta$ decays experiment JEF \cite{pro-JEF14,pro-JEF17}, and the $\eta/\eta^\prime$ factory REDTOP (which has been proposed as a part of the U.S.\ particle physics program~\cite{REDTOP:2022slw}) study fundamental symmetries in these systems and contribute to dark sector studies.

\subsection{ Fundamental Physics in Small Experiments}

The study of static properties of elementary particles (electric and magnetic moments) and  the fundamental symmetries ($C$, $P$, $T$, and their combinations, along with basic tests of Lorentz symmetry), all probe energies all through the Planck scale. The experiments addressing this physics  are of small to intermediate size and sometimes involve methods not traditionally considered high-energy physics. Nonetheless, in subjects such as $n\bar{n}$ oscillations, EDMs, and $0\nu2\beta$, the intellectual techniques, methods, and people frequently overlap, and  artificial divisions impede progress. We worked to identify ways to develop a synergistic physics program that exploits the expertise of scientists working in the high-energy physics community, the relevant AMO and nuclear physics to promote collaborative efforts that will make this research flourish.

Storage ring EDM experiments are an exciting opportunity. Proton storage ring experiments might reach $10^{-29} e\cdot cm$ in a decade, with deuterons achieving  similar levels between five and ten  years later \cite{Alexander:2022rmq,Alarcon:2022ero}. Like the FNAL $g {-} 2$ experiment, the technique calls for running at a ``magic momentum" (0.7 GeV/c for the proton).  The proponents have developed plans for a Brookhaven National Laboratory (BNL)-sited experiment, but the experiment could be sited elsewhere. 

Low-energy antimatter gravity tests and fifth force tests using muonium could be performed at the Advanced Muon Facility (AMF) facility described in Section~\ref{sec:clfv}.   These experiments are also synergistic with possible efforts for precision measurements of the muonium spectrum and searches for muonium-antimuonium oscillation. Studies for a low-energy muon facility are underway \cite{Johnstone_private}.


\subsection{Baryon and Lepton Number Violation}

The evolution of the Universe from its matter-antimatter symmetric state just after the Big Bang to its current asymmetric state requires the presence of one or both of baryon ($B$) or lepton ($L$) number-violating interactions. The Standard Model cannot produce the observed baryon-antibaryon asymmetry: the Higgs mass is too large to ensure that the electroweak phase transition is sufficiently first order and the size of CP-violation is too small.

We can search for new sources of baryogenesis in a set of experiments looking for possible baryon number violating processes, such as $B$-violating baryon decays and baryon-antibaryon oscillations. 
These searches are linked to the nature of neutrinos: if neutrinos  are Majorana particles and violate CP, there are well-motivated scenarios of leptogenesis that could explain the observed baryon asymmetry~\cite{doi:10.1146/annurev.nucl.55.090704.151558}.    

Experimental and theoretical research in $B/L$-number violation (such as proton decay, $n-\bar n$ oscillations, and searches for $0\nu 2\beta$ decays) is traditionally supported by nuclear physics. These efforts share many of the same intellectual problems, methods, and people with HEP. Unfortunately, this separation produces barriers that make it difficult for researchers at the intersection between these communities.  

Studies of neutrons are another avenue of opportunity for profound discoveries.  One particular opportunity of note is the study of $n\bar{n}$ oscillations.  The proposed NNBar experiment at the ESS could reach a limit of  $\tau_{n\bar{n}}\sim10^{9-10}\,$s \citep{Nesvizhevsky:2020vwx,Addazi:2020nlz,Gudkov:2019gro}. ``Mirror neutrons," $n\rightarrow n'$ oscillations, have been ruled out as an explanation of the discrepancy between neutron lifetimes measured with cold and ultracold neutrons, but improved sensitivity to this phenomenon is possible at the HIBEAM program at the ESS \citep{Addazi:2020nlz}.  

\subsection{Charged Lepton Flavor Violation} \label{sec:clfv}

Charged lepton flavor violating/violation (CLFV) processes are interactions that do not conserve lepton family number. Neutrinos change their flavor through oscillations, expressed in the PMNS matrix; quarks change theirs through the weak interaction and the CKM matrix. The archetypal CLFV decay is $\mu \rightarrow e \gamma$ (with no emitted neutrinos), compared to the normal weak decay of $\mu \rightarrow e \nu_{\mu} \bar{\nu}_e$.  We have never observed  CLFV transitions among muons, electrons, and taus. A fundamental question that arises from current observation is why mixing occurs in quarks,  whereas in the lepton sector, mixing occurs only among neutrino species.

Muons play a unique role in CLFV searches because we can make intense beams of muons and achieve the highest statistical sensitivity. Therefore it is important to explore initiatives that can push our sensitivity even further. 
In muon-to-electron conversion, $\mu^-N \rightarrow e^-N$, the Mu2e experiment at Fermilab will reach its goal of $R_{\mu e} = \Gamma(\mu^-N \rightarrow e^-N)/\Gamma(\mu^-N \rightarrow {\rm all~muon~captures}) < 8 \times 10^{-17}$ at 90\% CL by the end of the decade.  A factor of ten upgrade from Mu2e, Mu2e-II, needs R\&D effort to both improve the limit and change the physics probed with a different target nucleus $N$, most likely Ti \cite{Mu2e-II:2022blh}.  PIP-II at Fermilab could  not only provide profound improvements in this channel but also open intriguing possibilities for a comprehensive muon CLFV program.  The proposed AMF facility exploits the PIP-II beam and a pair of new rings.  It could improve limits on muon-to-electron conversion, $\mu^-N \rightarrow e^-N$, by at least one order of magnitude beyond Mu2e-II, and it would also enable the study the physics underlying a signal by varying the $Z$ of the nucleus $N$.  Using high-$Z$ materials such as Au is impossible  without PIP-II and the new rings.  Furthermore, the facility would allow us to examine the two muon rare decay modes $\mu \rightarrow e \gamma$ and $\mu \rightarrow 3e$ at rates far beyond those achievable at any other planned facility; here detector R\&D is essential to use the rates available at PIP-II.  AMF would also enable a dark matter experiment \cite{CGroup:2022tli}.  This program requires rebunching the PIP-II beam and a fixed-field alternating gradient synchrotron (FFA) \cite{Alekou:2013eta}. A proposal for such a  facility requires significant R\&D, and the R\&D required for targeting in a solenoid is closely related to that needed for the muon collider \cite{https://doi.org/10.48550/arxiv.2209.00142}.

Charged lepton flavor violating  $\tau$ decays are complementary to analogous $\mu$ decays, and they can be pursued in heavy-flavor experiments such as Belle II, LHCb, or a prospective tau-charm factory. GIM suppressions are smaller in the $\tau$ sector; thus we can set stringent limits for $\tau$-based CLFV with much smaller samples and limits of the order of  $10^{-9}$ to $10^{-10}$  are expected at the upcoming $e^+e^-$ experiments. CLFV in  Higgs decays can be studied at colliders by searching for $H \rightarrow \tau \mu$ and $ \rightarrow \tau e$~\cite{Harnik_2013}.


\subsection{Dark Sector at High Intensity}

The possibility of a dark sector neutral under Standard Model forces furnishes an attractive explanation for the existence of Dark Matter.  Dark sectors are a compelling new physics direction to explore in its own right, with potential relevance to fundamental questions as varied as neutrino masses, the hierarchy problem, and the Universe's matter-antimatter asymmetry.  Because dark sectors are generically weakly coupled to ordinary matter, and because they can naturally have MeV-to-GeV masses and respect the Standard Model's symmetries, they are only mildly constrained by high-energy collider data and precision atomic measurements. Intensity Frontier experiments offer unique and unprecedented access to experimental studies of the dark sector with possible access to specific dark matter candidates \cite{rpf6}. 

The continued exploitation of already existing large multipurpose detectors, especially Belle II and LHCb, to study dark-sector states is a crucial aspect of this physics program. In addition, dedicated efforts at high-intensity accelerators have started. The Dark Matter New Initiative (DMNI) report has given initial support to two experiments, and a promising set of experiments has been identified since the last P5 report \cite{osti1659757}. A broader, coordinated program is emerging from this initial work.  Dark-sector theory will also be critical. Advances in theory will both  address open problems in particle physics and cosmology, and  maximize the efficacy of the experimental program, where the track record of theorists pioneering new approaches is strong. A core consensus in the Frontier is that the U.S.\ should identify and pursue a well-planned portfolio of these experiments. 

\subsection{Hadron Spectroscopy}

Hadron spectroscopy encompasses the experimental investigation of the pattern of hadron masses and quantum numbers and the elucidation of their nature with non-perturbative QCD calculations. New states discovered by LHCb, Belle/Belle II, BESIII, and other experiments do not always fit among conventional hadronic states. New exotic states, such as tetra- and pentaquarks, hybrids, etc., are added every year. A key question arising from the multitude of newly discovered states is whether they can fit in a unified description of  multiplets with a similar underlying structure.  Configurations such as hadron molecules, compact multiquark states, or even linear combinations of such components, are possible and require careful experimental measurements to disentangle \cite{https://doi.org/10.48550/arxiv.2207.14594}.


\section{Experimental Methods
\label{sec:exp-methods}}
The physics drivers described above need a variety of experimental approaches, featuring different beams and detector designs, as well as different scales of investment and duration, to come to fruition. Some of the experiments considered have a solid data-taking record and are envisaging staged upgrades in different phases of development. Others are about to take data and are considering significant updates. Finally, new concepts are proposed and are currently in an early but promising stage.

\begin{figure}
    \centering
    \includegraphics[width=\textwidth,trim={340 210 260 330},clip=true]{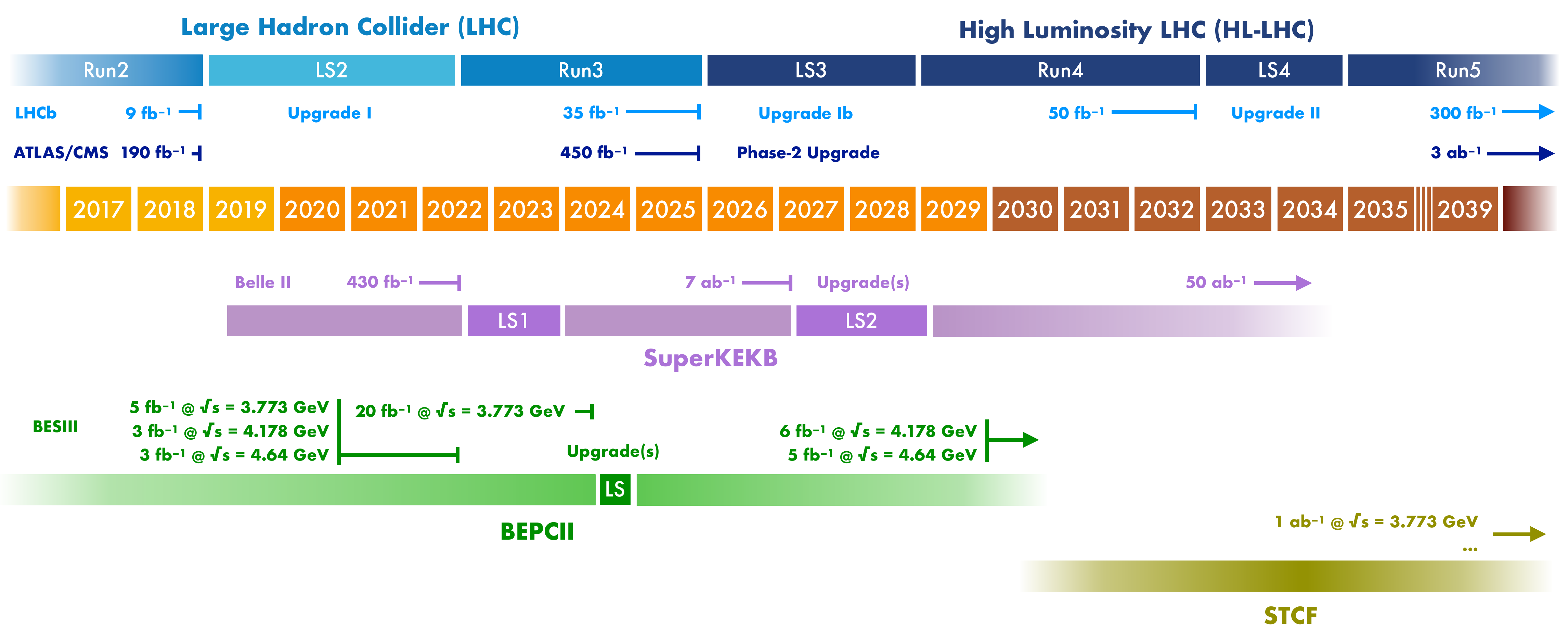}
    \caption{Timeline for the proposed experimental program in heavy flavor physics.\label{fig:tg1}}
\end{figure}

We begin our survey with high-precision heavy-flavor experiments; such experiments are essential to our physics program. Their timeline is summarized in Fig.~\ref{fig:tg1}.

Two of them encompass a vast program of measurements that use beauty and charm decays as tools for discovery: LHCb, the first experiment optimized to pursue heavy flavor decays at the LHC, and Belle II, operating at the SuperKEKB accelerator. They are both state-of-the-art detectors, featuring excellent tracking, particle identification, and high-acceptance electromagnetic calorimeters. LHCb operates at the LHC,  exploiting the higher $b$-production cross section and higher boost that allows superb vertex resolution and proper time resolution of the order of 40 fs. In addition, a great variety of $b$ and $c$  hadron species are accessible. Belle II exploits the simplicity of the $e^+e^-$ initial state to implement kinematic constraints that are effective in the study of inclusive properties of the lighter $B$ mesons, as well as quantum coherence of the initial $B\bar{B}$ state.

The LHCb experiment collected 9\invfb of $pp$ collisions during Run  I and Run II of the LHC and is now commissioning  its first major upgrade during the early data-taking phase of Run III. The first data set has already demonstrated the impressive range of measurements within LHCb reach, from the discovery of CP violation in the charm sector~\cite{LHCb:2019hro} and in the $B_s^0$ system, major inputs to the study of lepton flavor universality, and the first complete analysis of the angular observables in $b\to s\ell^+\ell^-$. LHCb is already making strides in measurements of rare \kaon decays and is planning to pursue rare hyperon decays as well. Heavy baryon-antibaryon oscillations~\cite{LHCb:2017vth} may provide clues to baryogenesis~\cite{Aitken:2017wie}. The field of hadron spectroscopy is already blossoming, with the observation of nearly sixty exotic hadronic states of matter, including pentaquarks~\cite{Workman:2022ynf}. In addition, LHCb has already put in place a program of measurements that provide complementary information to the big general-purpose experiments (ATLAS and CMS) in the study of key electroweak parameters, such as the $W$ boson mass, as well as searches  for heavy Majorana neutrinos and $W$ lepton flavor-violating decays~\cite{lhcbW2022,Aaij_2014,Aaij_2015}.  The most innovative feature of the LHCb upgrade I, currently being commissioned,  is the implementation of a software trigger that processes  data in real time~\cite{CERN-LHCC-2014-016}, 
rather than a 
simple first-level filtering scheme based on hardware thresholds.  This strategy~\cite{Aaij:2019zbu} allows the experiment to adjust the triggering algorithm to the evolving knowledge of these decays and to augment its efficiency for hadronic channels. This feature, combined with an increase in the maximum instantaneous luminosity the experiment can process, allows significant improvement in the measurement of  key quantities. The goal of LHCb upgrade I is to collect 50 \invfb. A subsequent major update is planned for installation in the time frame of CERN LS4 ($\approx$ 2031) and is currently incorporated in the LHC planning~\cite{LHCbCollaboration:2806113}. The goal is to maintain or improve the detector performance while taking data at an instantaneous luminosity of $1.5\times 10^{34}{\rm cm}^{-2}{\rm s}^{-1}$ and accumulate a data sample of 300 \invfb during HL-LHC operation. This requires operation at a much higher pile-up rate ($\sim$ 40 interactions per crossing). Thus a redesign of major detector components is needed, with special consideration given to the inclusion of precise timing (a few tens of ps) in the detector and associated electronics design. This feature makes possible the association of tracks and showers to specific primary vertices and significantly aids in 
pattern recognition. 

The Belle II experiment has been taking data at the asymmetric SuperKEKB $e^+e^-$ collider since 2019 and has collected $\sim 430 \invfb$ in the data-taking period ending with the Long Shutdown I of Summer 2022, roughly corresponding the sample collected by \babar.
While the current data-taking does not represent a significant increase in the statistical accuracy of the  Belle data set, novel techniques to improve the experiment's sensitivity have been implemented, with promising results illustrated by the $B\to K\nu\bar{\nu}$ mode~\cite{Belle-II:2021rof}.
The experiment aims to collect 50~$\invab$ by the mid-2030s.~\cite{Belle-II:2022cgf}. To achieve this goal, SuperKEKB needs to reach a peak luminosity of $6.5\times 10^{35}~{\rm cm}^{-2}{\rm s}^{-1}$ through upgrades planned for the Long Shutdown 2, currently scheduled for 2027--2028~\cite{Forti:2022mti}. An international task force has been formed to advise to SuperKEKB on the possible upgrade options, including a redesign of the interaction region and the final focus system. Long Shutdown 2 allows upgrading subsystems of the Belle II detector as well. In addition, a possible upgrade of the SuperKEKB machine to feature beam polarization would allow precision electroweak and precision $\tau$ physics measurements~\cite{https://doi.org/10.48550/arxiv.2208.05403}.

A similar experimental approach is used by the BES III experiment at BEPCII and may be further explored at the proposed super tau-charm (STCF) factory in China. In addition, the two general purpose experiments at the LHC, ATLAS, and CMS, have a heavy flavor program competitive in final states including muons and in spectroscopy studies. In the longer term, the FCC-ee program can build upon the strength of colliders to pursue flavor physics opportunities relevant to the time scale of operation.

Table~\ref{tab:sensitivity} summarizes the evolution of the precision achievable for important measurement as different upgrades are implemented and larger samples are accumulated. As the statistical accuracy improves with higher luminosity, the control of systematic effects and theoretical uncertainties becomes increasingly important. The complementarity in experimental approaches between $e^+e^-$ high luminosity collider experiments and experiments taking data at hadron machines is essential to the success of this physics program. LHCb and Belle II provide complementary information on key observables in \Bd and \Bp decays, as well as in hadron spectroscopy, searches for charged lepton flavor violation in $\tau$ decays and charged lepton flavor violation in $b$-hadron decays, and in the exploration of the dark sector.  Belle II has an easier path to inclusive measurements and exploits the simplicity of the initial state. LHCb and its upgrades have a higher production cross section and a richer array of $b$-flavored hadrons to study. Fundamental tests such as lepton universality violation or precision determinations of the CKM parameters $|V_{ub}|$ and $|V_{cb}|$ are prime examples of such a synergy, where a diverse array of methods and measurements may help in pinning down long-overdue unresolved tensions. Lastly we point out that study  of rare and forbidden $\tau$ decays, accessible in these experiments, is complementary to the several dedicated experiments searching for rare $\mu$ decays described below~\cite{https://doi.org/10.48550/arxiv.2203.14919}.

\begin{table}[ht]
\centering
\resizebox{\textwidth}{!}{
\begin{tabular}{lccccccccc}
\toprule
Observable & Current & \multicolumn{2}{c}{Belle II} & \multicolumn{2}{c}{LHCb} & ATLAS & CMS & BESIII & STCF\\
 & best & 50\invab & 250\invab & 50\invfb & 300\invfb & 3\invab & 3\invab & 20\invfb $(*)$ &  1\invab $(*)$\\
\midrule
\bf{Lepton-flavor-universality tests} \\
$R_K(1<q^2<6\ \gev ^2)$ & 0.044~\cite{LHCb:2021trn} & 0.036 & 0.016 & 0.017 & 0.007 \\
$R_{K^*}(1<q^2<6\ \gev ^2)$ & 0.12~\cite{LHCb:2017avl} & 0.032 & 0.014 & 0.022 & 0.009 \\
$R(D)$ & 0.037~\cite{Belle:2019rba} & 0.008 & $<0.003$ &na &na \\
$R(D^*)$ & 0.018~\cite{Belle:2019rba} & 0.0045 & $<0.003$ & 0.005 & 0.002 \\
\midrule
\bf{Rare decays} \\
${\cal B}(\Bs\to\mu^+\mu^-)$ [$10^{-9}$] & 0.46~\cite{LHCb:2021vsc,LHCb:2021awg} & & &na & 0.16 & 0.46--0.55 & 0.39\\
$\BF(\Bz\to\mu^+\mu^-)/\BF(\Bs\to\mu^+\mu^-)$ & 0.69~\cite{LHCb:2021vsc,LHCb:2021awg} & & & 0.27 & 0.11 &na & 0.21 \\
$\BF(\Bz \to K^{*0} \tau^+\tau^-)$ UL [$10^{-3}$] & 2.0~\cite{BaBar:2016wgb,Belle:2021ndr} & 0.5 &na \\
$\BF/\BF_{\rm SM}(B^+ \to K^+ \nu\bar{\nu})$ & 1.4~\cite{BaBar:2013npw,Belle:2017oht} & 0.08--0.11 &na \\
$\BF(B\to X_s\gamma)$ & 10\%~\cite{BaBar:2012fqh,Belle:2014nmp} & 2--4\% &na \\
\midrule
\bf{CKM tests and CP violation} \\
$\alpha$ & 5\degrees~\cite{BaBar:2014omp} & 0.6\degrees & 0.3\degrees \\
$\sin{2\beta}(\Bz\to\jpsi\KS)$ & 0.029~\cite{Belle:2012paq} & 0.005 & 0.002 & 0.006 & 0.003 \\
$\gamma$ & 4\degrees~\cite{LHCb:2021dcr} & 1.5\degrees & 0.8\degrees & 1\degrees & 0.35\degrees & & & $0.4\degrees\,(\dagger)$ & $<0.1\degrees\,(\dagger)$\\

$\phi_s(\Bs\to\jpsi\phi)$ & 32\mrad~\cite{LHCb:2019nin} & & & 10\mrad & 4\mrad & 4--9\mrad & 5--6\mrad \\
$|V_{ub}|(\Bz\to\pi^-\ell^+\nu)$ & 5\%~\cite{BaBar:2010efp,Belle:2010hep} & 2\% & $<1\%$ &na &na \\
$|V_{ub}|/|V_{cb}|(\Lb\to p\mu^-\nub)$ & 6\%~\cite{LHCb:2015eia} & & & 2\% & 1\% \\
$f_{D^+}|V_{cd}|(D^+\to\mu^+\nu)$ & 2.6\%~\cite{BESIII:2013iro} & 1.4\% &na & & & & & 1.0\% & 0.15\%\\
$\SCP(\Bz\to\eta^\prime\KS)$ & 0.08~\cite{BaBar:2008ucf,Belle:2014atq} & 0.015 & 0.007 &na &na \\
$\ACP(\Bz\to\KS\pi^0)$ & 0.15~\cite{BaBar:2008ucf,Belle:2008kbm} & 0.025 & 0.018 &na &na \\
$\ACP(D^+\to\pi^+\pi^0)$ & $11\times10^{-3}$~\cite{LHCb:2021rou} & $1.7\times10^{-3}$ &na &na &na & & &na &na\\
$\Delta x(\Dz\to\KS\pi^+\pi^-)$ & $18\times10^{-5}$~\cite{LHCb:2021ykz} &na &na & $4.1\times10^{-5}$ & $1.6\times10^{-5}$ & & & & \\
$A_\Gamma(\Dz\to K^+K^-,\pi^+\pi^-)$ & $11\times10^{-5}$~\cite{LHCb:2021vmn} &na &na & $3.2\times10^{-5}$ & $1.2\times10^{-5}$ & & & & \\
\bottomrule
\end{tabular}}
\caption{Projected uncertainties (or 90\% CL upper limits) in several key heavy-flavor observables over the next two decades. A missing entry means that the observable cannot be measured; the abbreviation {\rm na} means that, although the observable can be measured, the projected uncertainty is not available. Projections are taken from Refs.~\cite{Belle-II:2022cgf,Forti:2022mti,Belle-II:2018jsg} (Belle~II), Refs.~\cite{LHCb:2018roe,LHCbCollaboration:2806113} (LHCb), Ref.~\cite{ATL-PHYS-PUB-2022-018} (ATLAS and CMS), Refs.~\cite{BESIII:2020nme,Cheng:2022tog} (BESIII and STCF). $(*)$ Integrated luminosity at $\sqrt{s}=3.773$. $(\dagger)$ Projected uncertainties on $\gamma$ resulting from BESIII/STCF measurements of the $D$ strong-phase differences, which will contribute as external inputs to the Belle~II and LHCb measurements.\label{tab:sensitivity}}
\end{table}

The experimental study of kaon decays has played a crucial role in flavor physics as a tool for discovery, and exciting prospects are on the horizon. The ultra-rare $\Kp\to\pip\nu\bar{\nu}$ and $\KL\to\piz\nu\bar{\nu}$ decays with SM branching ratios of ${\cal O}(10^{-11})$ are theoretically well-understood and complement the experiments in the $B$ sector~\cite{Aebischer:2022vky} in the exploration of indirect evidences for new physics. A vibrant experimental program aiming primarily at the measurement of these decays is currently ongoing in Europe (NA62) and Japan (KOTO); the NA62 experiment has recently reported  evidence of the $\Kp\to\pip\nu\bar{\nu}$ decay with a $3.4\sigma$ significance~\cite{NA62:2021zjw}. Beyond the flagship modes, the two experiments pursue a broad program of rare kaon decay measurements~\cite{Cirigliano:2011ny}, tests of SM symmetries and CKM unitarity, and searches for hidden sectors~\cite{Goudzovski:2022vbt}, in a complementary way to other flavor experiments. The challenges of kaon experiments lie in the operation at GHz beam rates required for the collection of samples of ${\cal O}(10^{13})$ decays, precision timing (to sub-100~ps resolution), and background suppression. Both the NA62 and KOTO experiments are implementing significant upgrades and envisage long-term programs~\cite{NA62:2020upd}.

An experiment to probe lepton universality violation in BR$(\pi \rightarrow e \nu)$/BR($\pi\rightarrow \mu \nu$) and the rare pion beta decay  $\pi^+ \rightarrow \pi^0 e^+ \nu_e (\gamma) $ decay has been recently approved at the Paul Scherrer Institute~\cite{PIONEER:2022yag}. The experiment is currently in the R\&D phase, where technological options that allow optimum separation between signals and potential backgrounds are being considered. The targeted measurements are connected to recently reported anomalies~\cite{PIONEER:2022yag}. The technologies being explored are active, segmented targets using AC coupled low gain avalanche detectors (AC-LGADs) to implement a 4-dimensional hit reconstruction, surrounded by a fast, deep, and high-resolution calorimeter~\cite{https://doi.org/10.48550/arxiv.2209.14111}. Potential technologies are a LXe calorimeter or a hybrid LYSO + CsI crystal combination. In addition, state-of-the-art trigger and data-acquisition systems must be implemented.

Another approach to exploit large samples of light mesons as a tool for discovery is the use of large samples of $\eta$ and $\etapr$ mesons, as in the upcoming JEF experiment~\cite{pro-JEF14,pro-JEF17} and the proposed REDTOP experiment~\cite{REDTOP:2022slw}. Their main goals are to search for the violation of fundamental symmetries or lepton flavor universality, and possibly to study rare decays of these mesons as a probe of particles that act as portals to a possible dark sector. The proposed REDTOP detector is a 4$\pi$ detector surrounding a Li target. The expected rate is 1 GHz of inelastic collisions, with 5.1$\times 10^6$ $\eta$'s produced. The $\eta$ and \etapr mesons are produced almost at rest; the minimization of the material budget in the detector is thus a primary consideration. In addition, the experiment has to cope with an inelastic event rate of  $\approx 7\times 10^8$ Hz; thus the trigger and data processing  pose significant challenges. Although several detector choices  and the facility hosting the experiment still have to be finalized, the experiment aims to take data starting within the next decade. 

\begin{figure}[ht]
\begin{center}
\includegraphics[width=0.7\textwidth]{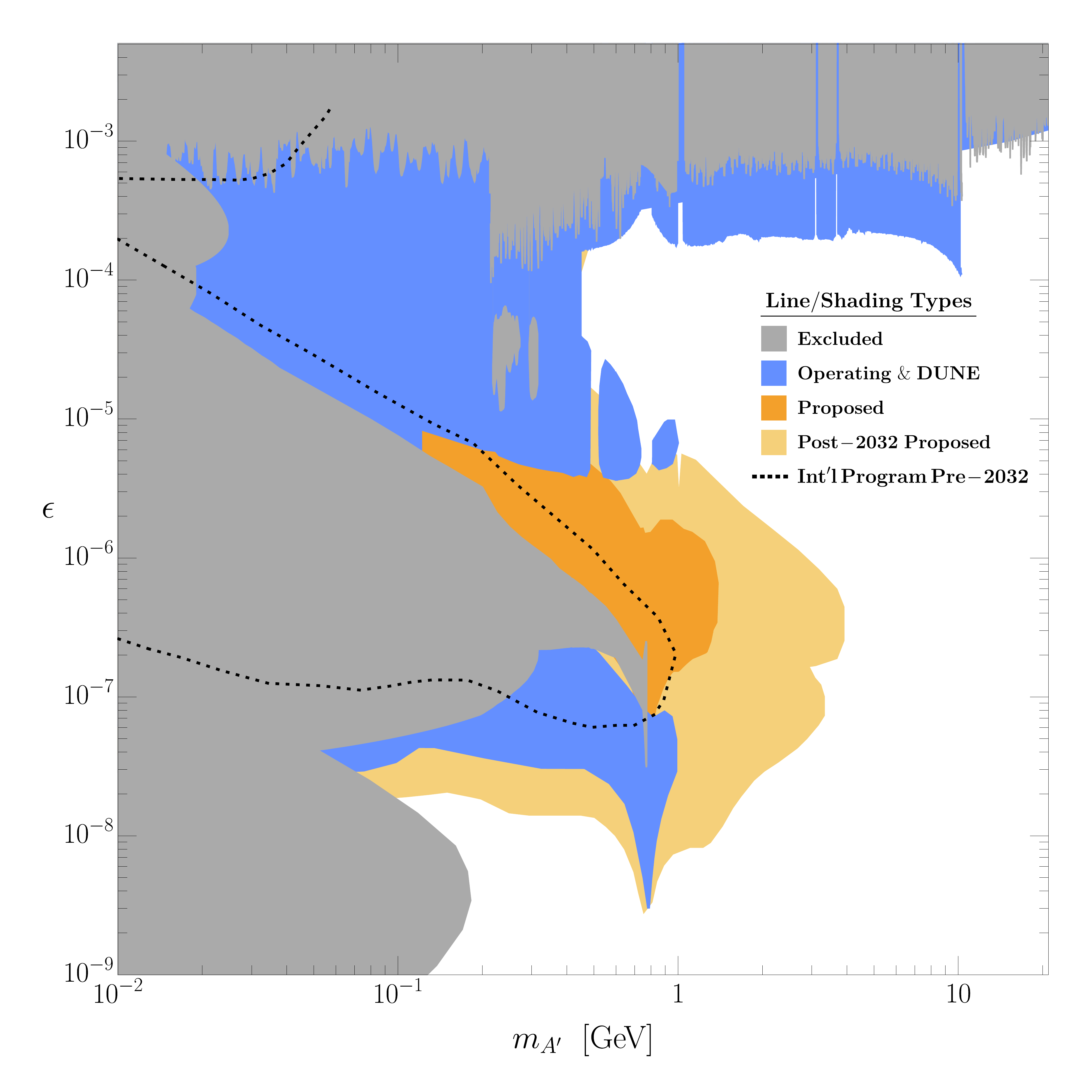}
\end{center}
\caption{ 
Prospects on near-term sensitivities and future opportunities to search for visibly decaying dark photons interacting through the vector portal. Exclusion limits are shown in terms of the dark photon mass $(m_{A'})$ and kinetic mixing  $(\epsilon)$ parameters~\cite{https://doi.org/10.48550/arxiv.2207.06905}.  This parameter space is compatible with secluded heavier thermal dark matter~\cite{Pospelov_2008}. 
Constraints from past experiments (gray  regions) and projected sensitivities from operating and fully funded experiments and DUNE (blue regions),  and other proposed near-term (pre-2032) experiments both based in the U.S.\ or with strong U.S.\ leadership (orange region) are shown. 
Primarily international projects are shown as a dashed line, and 
proposed experiments farther into the future are displayed in light yellow. 
This plot is for only one portal; similar curves for other cases are given in the subgroup report. Figure and caption adapted from ~\cite{Gori:2022vri}.
}
\label{rf6-fig:phys-BI2-aprime-cartoon}
\end{figure}

A broad program of dark sector studies at intensity machines is a pillar of our proposed research program. The nature of dark matter remains elusive, and the ``WIMP miracle'' did not materialize as predicted. A well-motivated possibility is that the dark matter is composed of particles that are  part of a dark sector similar in structure and perhaps complexity to the SM. Dark matter particles might interact with other dark matter particles via dark forces mediated by new gauge bosons and with ordinary matter through the exchange of dark forces.  These new forces may have structures similar to the known forces~\cite{Krolikowski:2008qa}. For example, dark matter particles could interact via a dark force similar to the electromagnetic force felt by ordinary matter. The dark photon $A^\prime$ that mediates this force can have a weak coupling to the standard electromagnetic current through the kinetic term $\epsilon F^{\mu\nu}F^\prime_{\mu\nu}/2$, where $\epsilon$ characterizes the strength of the mixing. Figure~\ref{rf6-fig:phys-BI2-aprime-cartoon} illustrates the parameter space that could be probed in the near and intermediate time scales.

\begin{figure}[ht]
    \includegraphics[width=6.0in]{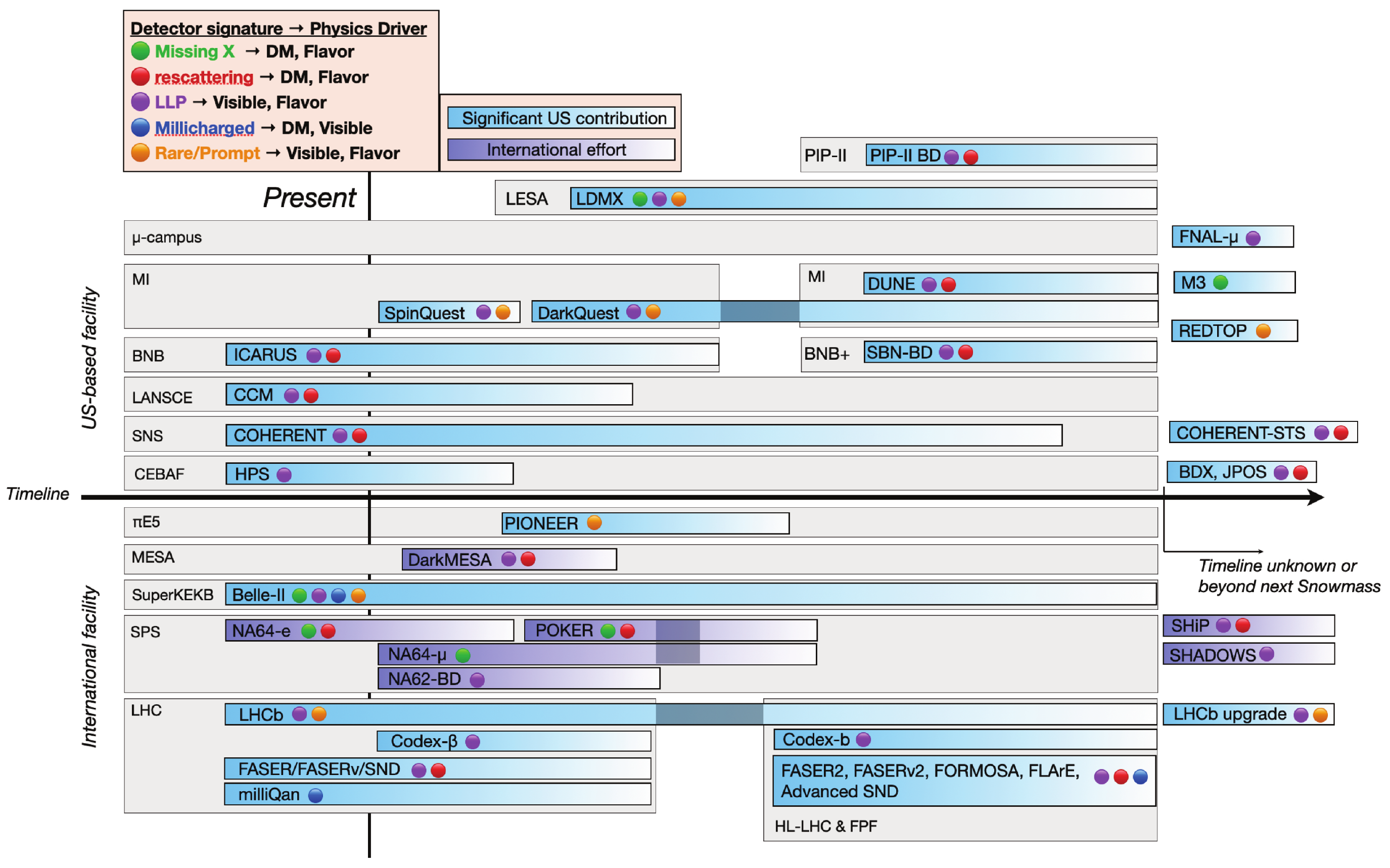}
    \caption{Timeline for the proposed experimental program to study the dark sector at high-intensity beams and  machines (from \cite{Ilten:2022lfq}).}
    \label{fig:tg6}
\end{figure}

 Experiments using intense beams offer unique and unprecedented access to the dark sector with possible discoveries of  specific dark matter candidates.
The characteristics of low mass and weak couplings of these exotic particles make Intensity Frontier experiments particularly suited to this kind of physics. An overview of the many experiments proposed to explore the dark sector in intense beams and machines is presented in Fig.~\ref{fig:tg6}.  A key aspect of this physics program is the continued exploitation of already existing large multipurpose detectors, especially Belle-II and LHCb, to study dark sector states. In addition, dedicated efforts at high-intensity accelerators have started. A Basic Research Needs (BRN) workshop on Dark Matter New Initiatives identified the study of dark matter particles and associated forces below the proton mass, leveraging DOE accelerators that produce beams of energetic particles~\cite{osti1659757}. This workshop resulted in support for two experiments. One of them,  Coherent CAPTAIN-Mills (CCM)~\cite{VandeWater:2022qot}, is a  beam dump experiment  employing the 800-MeV proton beam at the Los Alamos Science Center (LANSCE).  It uses a large liquid argon fast detector and timing synchronization with the source to reject background. An improved detector (CCM200) with 200 inner PMTs and a purification system to achieve lower thresholds began taking beam and calibration data in 2021 and has started operation. The second one is LDMX~\cite{LDMX:2018cma}, an electron fixed-target experiment that searches for sub-GeV dark matter via the missing momentum technique at the Linac to End Station A  (LESA) facility at SLAC. It utilizes a W or Al target surrounded by a hermetic detector and searches for a massive particle that can escape without depositing significant energy in the detector. It has received pre-project funds and needs further support to become operational. In addition, a promising set of experiments has been identified since the last P5 report~\cite{Gori:2022vri,Ilten:2022lfq}. Dark-sector theory will also be critical both to continue developing dark-sector models to address open problems in particle physics and cosmology and to maximize the efficacy of the experimental program, where the track record of theorists pioneering new approaches is strong. 

\begin{figure}[ht]
\includegraphics[width=1.0\textwidth]{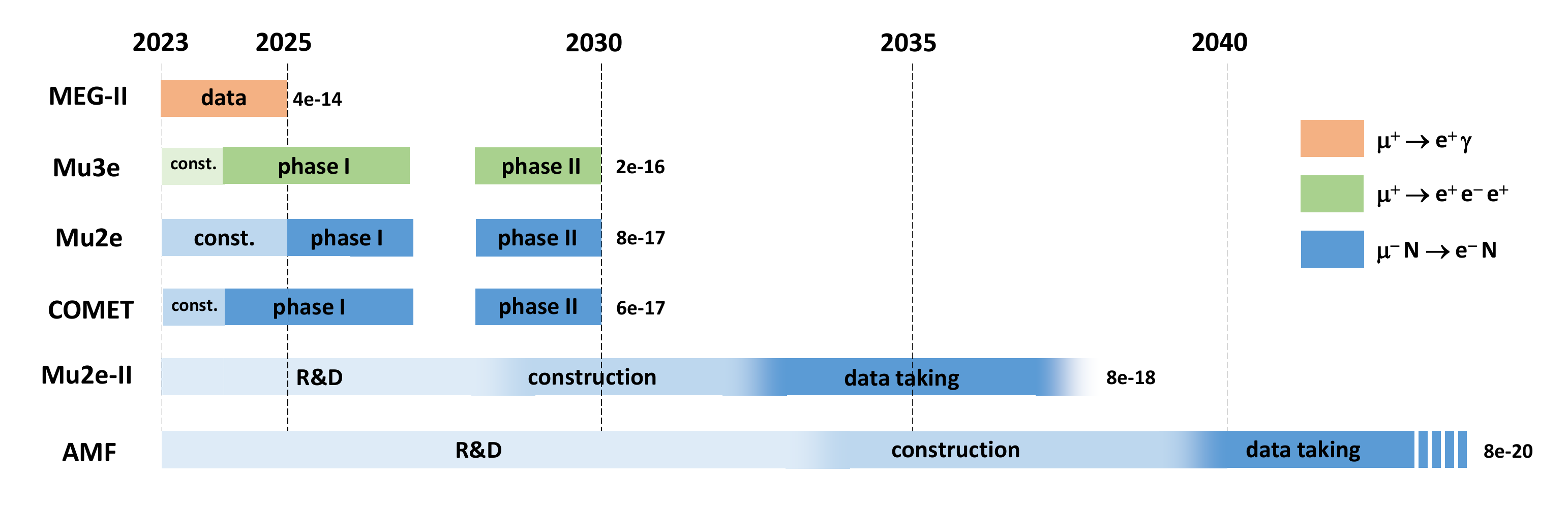}
\caption{\label{fig:clfvTimeline} Timeline for muon-based charged lepton flavor violation experiments discussed in this report. Approximate expected dates are shown. The number to the right of the timeline is the expected final 90\% CL.}
\end{figure}

A vibrant program of new physics searches in rare $\mu$ decays is currently ongoing in Europe (PSI) and Japan (J-PARC) and is about to start in the U.S.\ (Mu2e).  An approximate timeline is presented in Fig.~\ref{fig:clfvTimeline}.
The PSI program currently consists of  the MEG-II 
experiment, searching for the decay $\mu^+ \rightarrow
e^+ \gamma$, and Mu3e, searching for $\mu^+ 
\rightarrow  e^+e^+e^-$ in muon decay. MEG-II is an 
upgrade of the MEG experiment and will collect its 
data starting in 2022~\cite{sym13091591}. Mu3e Phase 
I will take data by the middle of the decade. The HiMB upgrade at PSI will enable about an order of magnitude improvement in the discovery potential for Phase-II of the Mu3e experiment starting around 2028~\cite{Hesketh:2022wgw, https://doi.org/10.48550/arxiv.2111.05788}. However, a new experimental concept will be required for MEG to use the improved HiMB beam~\cite{Renga_2019}.

The third muon 
CLFV process, coherent muon-to-electron 
conversion $\mu^- N \rightarrow e^- N$, will be pursued by
 the Mu2e  and COMET  experiments beginning 
around 2025, at FNAL and J-PARC respectively.  Assuming ${\cal O}(1)$ couplings in the EFT Lagrangian for new physics in these decays,  the
experiments described here will probe mass scales for 
dimension-6 magnetic dipole operators  at a 
few $\times 10^3$ TeV/c$^2$. Muon-to-electron conversion   
and $\mu \rightarrow 3e$ are sensitive to more 
operators than $\mu \rightarrow e \gamma$.  Both COMET and Mu2e can reach mass scales 
of ${\mathcal O}(10^4)$ TeV/c$^2$ in
dimension-6 contact terms.  We also note that
\mutoe and, to a lesser extent, COMET can probe the $\Delta L =2$ process $\mu^- N\rightarrow e^+N^{\prime}$, which is sensitive to new particles such as leptoquarks and is related to neutrinoless double-beta decay through the Schechter-Valle theorem~\cite{Schechter:1980gr}.

The U.S.\ program focuses on the muon conversion process $\mu^- N \rightarrow e^- N$. The possibility of utilizing different targets may elucidate the nature of the new physics possibly unveiled through studying  these decays. In addition, the combination of results obtained studying these decays with results from experiments studying $\mu^+ \rightarrow e^+ \gamma$  and $\mu^+ \rightarrow e^+e^+e^-$ transitions may pin down the  source of new physics. The Mu2e experiment, recommended by the previous P5 panel, is currently being constructed.  Mu2e will  have its first run at lower beam intensity in 2025, just before the LBNF/PIP-II shutdown. Its goal is to improve by about four orders of magnitude upon the limit on the parameter $R_{\mu e}$ obtained by the SINDRUM II experiment at PSI~\cite{SINDRUMII:2006dvw}. The key features of the Mu2e experiment are a high-resolution tracking system and a calorimeter placed inside a solenoid. The muon transport line is based on curved solenoids to shield the detector from the direct line of sight of the production target and select negatively charged muons.  In parallel, the Mu2e experiment is investigating an upgrade (Mu2e-II at PIP-II) that can  enhance their sensitivity by about a factor of ten, beginning with  an increase of the beam power from 8 kW to 100 kW~\cite{Mu2e-II:2022blh}. The detector and beamline will then have to cope with increased radiation and backgrounds. Studies of production solenoid modifications, along with a  reduction of material in the tracking relative to Mu2e, along with novel and faster calorimeter materials, are key goals of the proposed R\&D~\cite{Mu2e-II:2022blh}. In the longer term the suggested AMF facility will require significant R\&D; some of these efforts, in particular high-power targeting inside a solenoid, are related to R\&D for the muon collider~\cite{CGroup:2022tli}.

\begin{figure}[ht]
\includegraphics[width=1.0\textwidth]{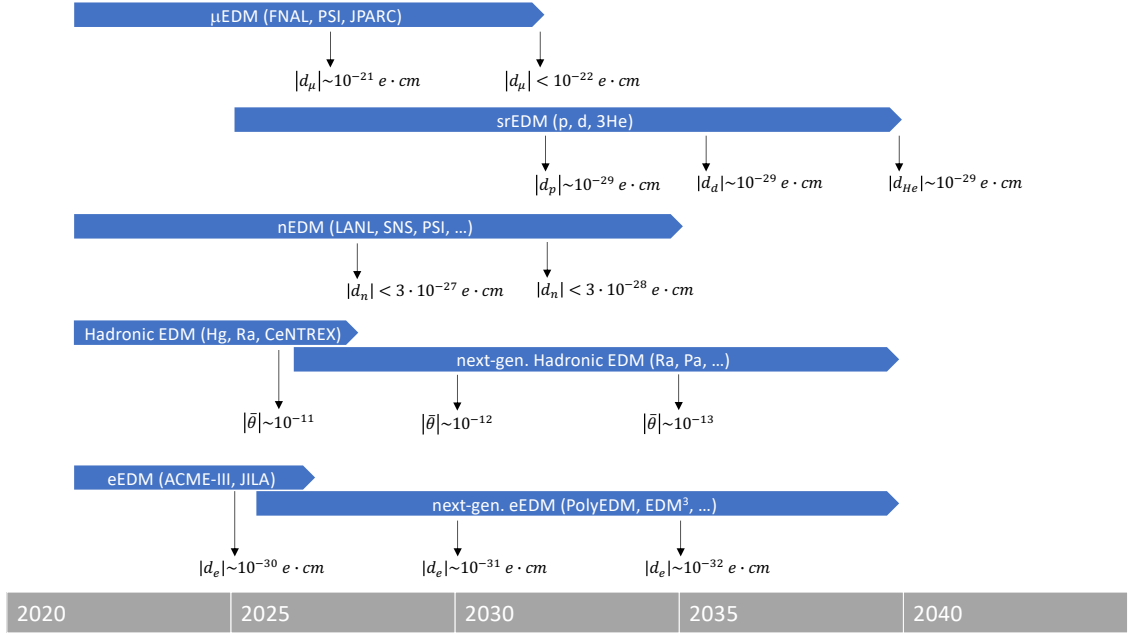}\vglue -1.0in
\caption{\label{fig:edmTimeline}Timelines for the major current and planned EDM searches.  Their measurement goals are indicated by the black arrows, based on the current plans of the various groups. The Storage Ring proton EDM experiments highlighted in this report are labeled ``srEDM." }
\end{figure}

Flavor-conserving processes such as the muon anomalous magnetic moment, or electric dipole moments (EDMs) of any system (electron, muon, neutron, proton, atom, or molecule),  would represent an exciting manifestation of new physics. As a part of the Snowmass process, the current landscape of proposed electrical dipole moment measurements has been examined~\cite{Alarcon:2022ero}. The search for the neutron EDM started by Ramsey and Purcell~\cite{Smith:1957ht} has reached a precision of $(0.00\pm 1.1|_{\rm stat}\pm 0.2|_{\rm sys})\times 10^{-26} e\cdot{\rm cm}$~\cite{Abel:2020pzs}. 
Atoms and molecules have been increasingly powerful EDM probes, exploring CP violation through EDMs, and several searches leveraging techniques developed in AMO physics have been proposed in order to increase the sensitivity by a factor of 10 in the near-term time scale, ultimately aspiring to achieve a four-to-six order of magnitude improvements in a 15--20-year time scale, exploiting progress in quantum science techniques, and availability of exotic nuclei, made possible by advances in nuclear physics.  A new approach is the proposed EDM measurement at a storage ring~\cite{Alexander:2022rmq,Alarcon:2022ero}. This technique, inspired by the method used to measure the anomalous magnetic moment of the muon, has the goal of reaching a proton EDM sensitivity of $10^{-29}e\cdot {\rm cm}$ while probing other  fundamental quantities, such as CPV in the Higgs sector, and improving current limits on $\theta _{\rm QCD}$ by three orders of magnitude. A timeline for EDM experiments is given in Fig.~\ref{fig:edmTimeline}.

Finally, violations of discrete conservation laws, such as baryon number or lepton number, or $B$--$L$, which are preserved by the SM, are a pillar of the experimental program described here. Their study supports our understanding of the baryon asymmetry in the universe, with the consequent disappearance of antimatter. Thus the search for proton decay with an increased sensitivity, an integral component of the DUNE and Hyper-Kamiokande physics programs, is closely connected with the themes and big questions  we are investigating in this Frontier. In addition, the nature of the neutrino species, and whether they are Dirac, Majorana, or a mixture of these two, is a  fundamental question  in particle physics. We have shown how some of the experiments described so far can investigate heavy Majorana neutrinos and baryon oscillations. A landmark set of measurements that can elucidate whether light neutrinos are Majorana or Dirac particles is based on studies of neutrinoless double-beta  decays. Because of their nature, they are an investment of the experimental nuclear physics community, but there is an extensive and profound synergy with the high-energy physics community that is equally interested in this fundamental question. In particular, the ``multi Tonne" scale experiments that plan to reach a sensitivity of $T_{1/2}=10^{28}$ years naturally  lend themselves to shared HEP and NP efforts. Achieving this goal will require optimizing the energy resolution of the detector, reducing backgrounds, and measuring residual backgrounds with high efficiency.  Only then can the experiment build a precise background model and provide a reliable estimate of the experiment's sensitivity. These are areas where experience in the high-energy physics community is crucial.  On the other hand, the  experimental methods and calculations required for studying a variety of isotopes  are the provinces of nuclear physics. Thus a fruiful collaboration between these two communities would greatly aid the successful implementation of this fundamental physics program.

\section{R\&D for the Rare and Precision Frontier}
\label{sec:rd}

Realizing the experiments described in this report \cite{https://doi.org/10.48550/arxiv.2209.00142} requires R\&D in both 
accelerators and instrumentation.  The muon program requires targeted accelerator  R\&D on three fronts.   The Accelerator Frontier AF5 report on ``Accelerators for Rare Processes and Physics Beyond Colliders" discusses many of these issues in more detail~\cite{https://doi.org/10.48550/arxiv.2209.06289}:
\begin{itemize}
\item A compressor ring to rebunch the PIP-II beam.  The ring should be designed to support both the muon program and a dark-matter beam dump 
experiment.  A particular challenge here is the kickers that need to operate at up to about 40 kHz \cite{CGroup:2022tli}.
\item a fixed-field alternating gradient synchrotron to serve as a muon storage ring.  A prototype has been constructed at Osaka \cite{Witte:2012zza}.
\item The experimental program requires targeting the PIP-II beam at $\geq 100$ kW inside a superconducting solenoid, with an ultimate goal of 1 MW of power, well-matched to PIP-II. Achieving 1 MW will require an upgrade to or replacement of the Fermilab Booster, enabling a wide range of physics \cite{arrington2022physics}.  The problem shares many aspects with some schemes suggested for  targeting for a muon collider; this program could then serve as a ``test bed" for muon collider R\&D \cite{https://doi.org/10.48550/arxiv.1901.06150, Gourlay:2022ktb,https://doi.org/10.48550/arxiv.2209.01318}. 
\end{itemize}

The experiments examined in this frontier can be grouped into four broad categories: experiments studying interesting quark transitions and $\tau$ lepton decays, experiments searching for lepton flavor violation in $\mu$ decays, experiments searching for EDMs, and dark sector at high-intensity experiments. We will now consider some of the instrumentation R\&D required for this program.

Table~\ref{tab:requirements} summarizes the major detector R\&D requirements of these experiments, organized in terms of solid-state tracking devices, gaseous tracking devices, calorimetry, VLSI electronics, trigger and data acquisition, and quantum sensors in the medium-term. 

A common theme in several of these experiments is the importance of precision timing information, with a resolution of the order of tens of picoseconds, to reduce combinatorial backgrounds associated with multiple interactions per crossing in high luminosity hadron machines or as a tool for out-of-time background suppression. 

Increasing the luminosity to reach higher sensitivities also implies unprecedented radiation fluence. For example, the LHCb upgrade 2 will withstand fluences up to $5\times 10^{16} {\rm n_{eq}/cm^2}$ during its nominal running time. Even in $e^+e^-$ high luminosity colliders such as Belle II, the ambitious upgrade planning makes  radiation hardness an important consideration.

\begin{table} [ht!]\small%
\caption{Detector specifications for representative RPF experiments (medium term goals)\label{tab:requirements}}
\begin{tabular}{ | p{15em} | p{3cm}| p{3cm} | p{5cm}|} 
\hline
 {Experimental approach}\hfill & Technology & Property & Requirement \\
\hline\hline
\multirow{3}{*}{Quark flavor experiments} \hfill 
~ & Solid  State Tracking Detectors & Time stamp  & 10--30ps/hit in the silicon pixel vertex detector  \\
&  & Radiation hardness & \hbox{fluences up to $5\times 10^{16} {\rm n_{eq}/cm^2}$}\\\cline{2-4}
 & Calorimetry & \hbox{Time stamp resolution} & 10-30 ps/shower\\\cline{2-4}
~~ & Trigger \& DAQ & Real time processing & 400-500 TB/sec\\
~~& ~~ & Optical links & Radiation-hard, fast, low-power and low-mass\\\hline
LFV experiments ($\mu)$ & Staw Tube Tracker & timing & 20 ps/track, low-mass, excellent momentum resolution \\\cline{2-4}
~~ & Calorimetry & Energy resolution & $<10\%/\sqrt{E}$, low cost\\
~~ & ~~ & timing & Shower time stamp $< 500$ ps, dose $>$ 900 KRad\\\hline
EDMs & Controlled preparation of many coherent particles & coherence times& $\tau\ge 1s$ or $N>>10^{12}$\\\cline{2-4}
~~& Laser locking, tuning and linewidth narrowing (many narrow-band lasers on target) & tunable narrow band &$<1$ MHz, $\lambda=2002-400{\rm nm}$\\ \hline
Dark~Sector (missing energy technique)& E\&M calorimetry & Energy resolution &~~\\
\hline
 Dark Sector (Beam Dump experiments) & photosensors & fast-timing photosensors  & ~~\\\hline
\end{tabular}
\end{table}

Operation at  ever-increasing luminosity and the correspondingly harsher environments will pose several challenges for the front-end electronics. The need to incorporate higher and higher complexity in devices that need to sustain higher radiation fluences
implies the use of a variety of technologies, such as CMOS devices with ever-smaller feature sizes and
possible alternatives to CMOS technologies. The effort to understand the transistor parameters of different technologies and to streamline the design to avoid prohibitive development costs is common to many of the projects described here. This is one of the most promising targets of a national R\&D program, where a multi-frontier collaborative efforts are established to develop the technologies needed to accomplish the broad scientific goals of the US high-energy physics community\cite{Artuso:2022qfq}.

On a longer time scale, all these requirements will apply but with more stringent specifications.  For example, the Mu2e-II experiment will require ``transparent tracking" with a reduction of the 15$\mu$m straw walls down to 8$\mu$m and faster and more radiation-hard crystal calorimeters. Quark flavor experiments implemented at a high energy and luminosity $e^+e^-$ collider will require a timing resolution at the level of 1 ps per track for $\pi p K$ separation. Radiation requirements will be pushed to $10^{18}{\rm n_{eq}/cm^2}$. The energy resolution and background suppression required by the dark sector experiments utilizing a missing energy technique are synergistic with some of the technology efforts described here. Experiments pursuing EDM measurements and other precision tests of fundamental laws are leveraging the progress in AMO techniques and quantum sensors to exploit a broader array of probes, a variety of molecules and different nuclei (some of them with extreme properties such as high angular momentum), and bigger samples with longer coherence times.  The combination of these efforts is building a foundation for an era of discovery~\cite{https://doi.org/10.48550/arxiv.2209.14111}.

\section{The Role of Theory  \label{sec:theory}}

\begin{figure}[t]
\centering
\includegraphics[width=0.5\textwidth]{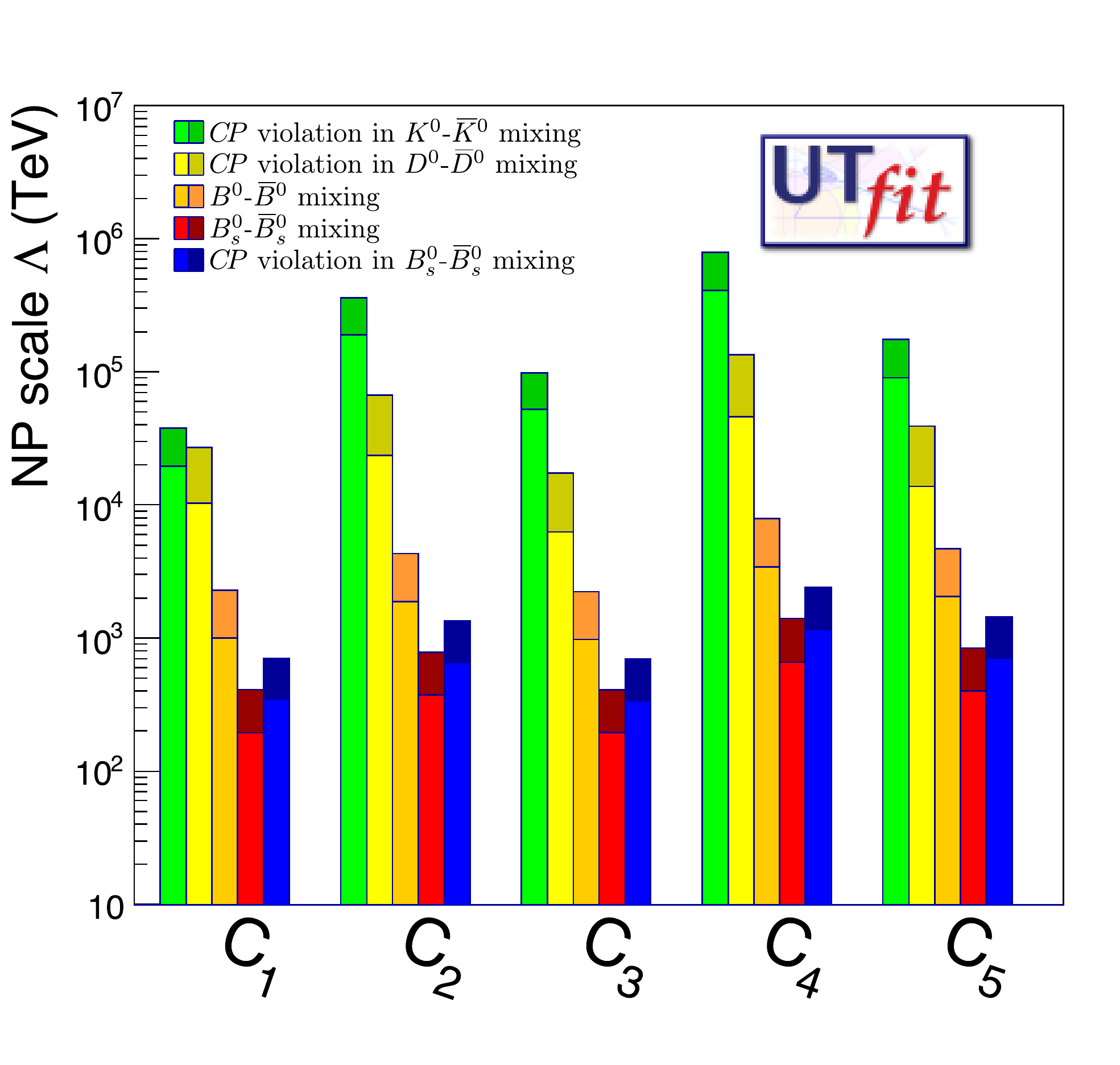}\hfil
\includegraphics[width=0.5\textwidth]{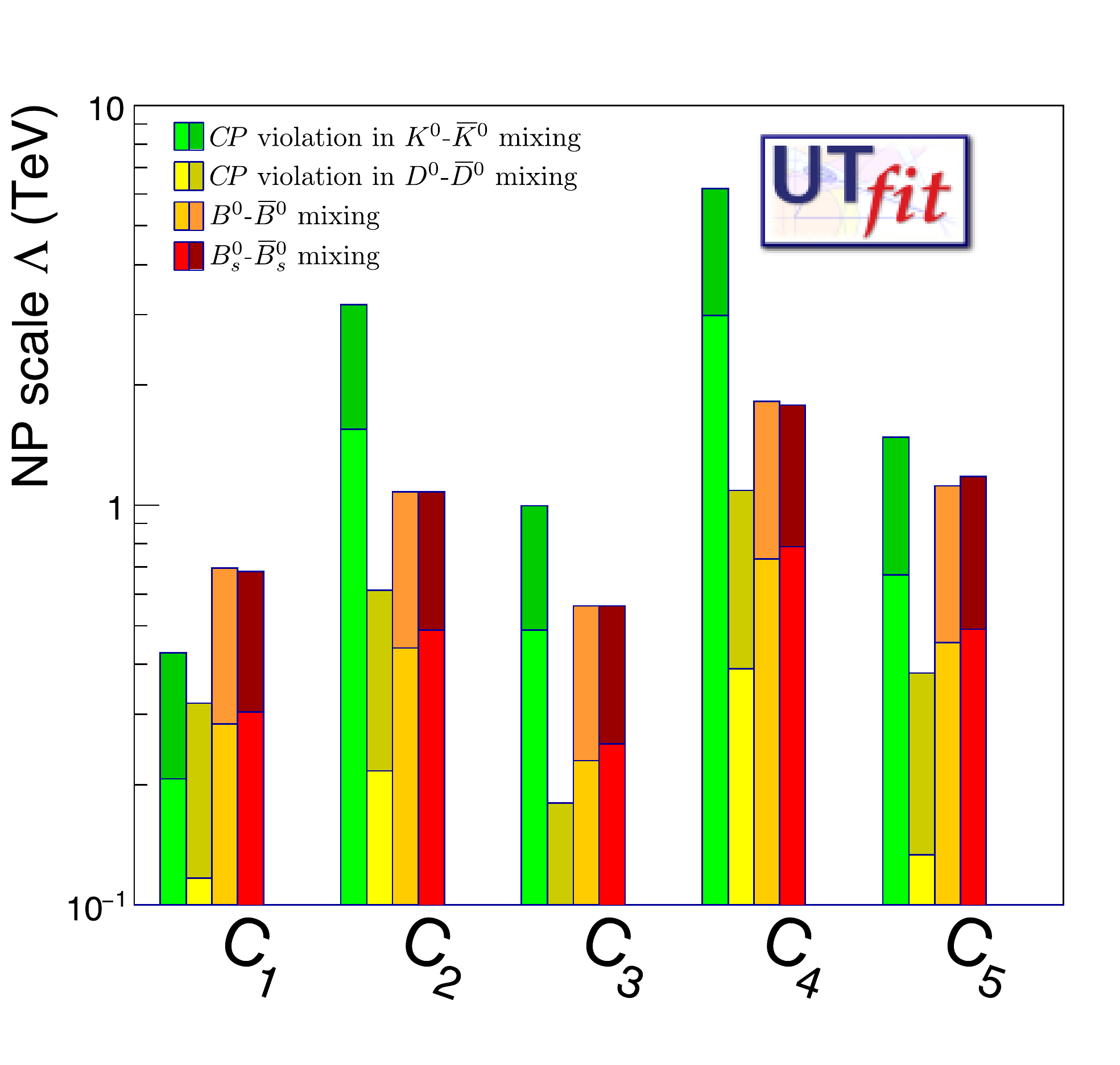}\\
\caption{Present (lighter) and future (darker) lower bounds at 95\% confidence level on the NP scale $\Lambda$ from $\Delta F=2$ transitions~\cite{UTfit:2007eik,Cerri:2018ypt}. The Wilson coefficients $C_i=F_iL_i/\Lambda^2$ ($i=1,...,5$) are the coupling of the NP dimension-six operators governing the $\Delta F=2$ transition: $Q_1^{q_iq_j}=(\bar{q}_{jL}^\alpha\gamma_\mu q_{iL}^\alpha)(\bar{q}_{jL}^\beta\gamma^\mu q_{iL}^\beta)$, $Q_2^{q_iq_j}=(\bar{q}_{jR}^\alpha q_{iL}^\alpha)(\bar{q}_{jR}^\beta q_{iL}^\beta)$, $Q_3^{q_iq_j}=(\bar{q}_{jR}^\alpha q_{iL}^\beta)(\bar{q}_{jR}^\beta q_{iL}^\alpha)$, $Q_4^{q_iq_j}=(\bar{q}_{jR}^\alpha q_{iL}^\alpha)(\bar{q}_{jL}^\beta q_{iR}^\beta)$, and $Q_5^{q_iq_j}=(\bar{q}_{jR}^\alpha q_{iL}^\beta)(\bar{q}_{jL}^\beta q_{iR}^\alpha)$ (with $\alpha$ and $\beta$ being color indices). On the left, NP is assumed to have arbitrary flavor structure ($F_i=1$) and to be strongly coupled with no loop suppression ($L_i=1$); on the right, NP is assumed to have minimal-flavor-violation couplings ($F_i=V_{\rm CKM}$) and to enter at one loop with weak coupling ($L_i=\alpha_2^2$, with $\alpha_2^2$ being the weak structure constant). The future bounds are based on expected sensitivities at Belle II (50\invab), BESIII, LHCb Upgrade II (300\invfb), and ATLAS/CMS (3\invab), and on improved theory inputs (from \cite{https://doi.org/10.48550/arxiv.2207.14594}).\label{fig:NPscales}}
\end{figure}

The primary strategies for indirect searches of New Physics particles proceed along three main directions: (1) studies of processes that are not allowed in the Standard Model, (2) studies of processes that are not allowed in the Standard Model at the tree level, and (3) studies of the processes that are allowed in the Standard Model. While virtual effects associated with new particles can probe energy scales beyond direct detection experiments (see, e.g., Fig.~\ref{fig:NPscales}), there are low-energy SM  contributions that can potentially mask or even overwhelm such effects. Thus, proper interpretation of experimental results is crucial in searching for New Physics phenomena. 

The role of theory in the unambiguous interpretation of the results of experiments and requirements for theoretical methods used in the analyses have significantly changed over the last thirty years, shifting from model estimates and quenched lattice QCD calculations to rigorous approaches with improvable computations of theoretical uncertainties \cite{Boughezal:2022cbl,Davoudi:2022bnl}. 
The theoretical interpretation of experimental data often relies on the computation of matrix elements of operators of some effective Lagrangian ${\cal L}$. The construction of those operators and the required calculations of matrix elements rely on a separation of the physical scales often present in a problem. The methods of Effective
Field Theories allow relating physics at different scales by performing matching and employing the running of the renormalization group \cite{Petrov:2016azi}. 

Theoretical methods applied to interpreting experimental data often depend on the assumption about the properties of NP particles, in particular on the energy scale at which NP degrees of freedom can propagate on-shell. The EFT techniques allow for incorporating this fact into the computations directly. 

For heavy NP, the masses of all NP particles are larger than the energy scales associated with an experiment. Explicit modeling of New Physics effects allows one to conjecture a fundamental mechanism that could have implications for the low- and medium-energy phenomena. Examples of such modeling include multi-Higgs or leptoquark models with flavor-changing neutral current interactions \cite{DiLuzio:2017vat} (see \cite{https://doi.org/10.48550/arxiv.2208.05403} for a recent review of the models) or models of Grand Unification that contain new particles mediating baryon- and lepton-number violating interactions \cite{Georgi:1974sy,FileviezPerez:2010gw} (see \cite{https://doi.org/10.48550/arxiv.2208.00010} for a  recent review of the models). Such explicit models are extensively discussed in the Topical Group reports   \cite{https://doi.org/10.48550/arxiv.2208.05403,https://doi.org/10.48550/arxiv.2208.05403,https://doi.org/10.48550/arxiv.2209.08041,https://doi.org/10.48550/arxiv.2209.00142,rpf6,https://doi.org/10.48550/arxiv.2207.14594}.

No on-shell propagation of NP particles is possible at energy scales below the masses of such new particles. In that case, the effects of NP can be approximated by a series of operators of increasing dimension multiplied by Wilson coefficients encoding the effects of NP particles:
\begin{equation}\label{lagrangianEFT}
{\cal L}_{\rm eff} = {\cal L}_{\rm SM } + \sum_{i,n} \frac{C_{i,n}(\mu)}{\Lambda^{n}} Q_i^{(n)}(\mu),
\end{equation}
where $\Lambda$ parameterizes the scale at which New Physics becomes operational and $C_{i,n}$ represents the Wilson coefficients, which are given by a combination of coupling constants and other factors. Each model of NP provides different combinations of factors. The EFT methods are uniquely suited to computations of indirect effects of new physics, as the construction of Eq.~(\ref{lagrangianEFT}) parameterizes {\it all} BSM models that can contribute to a given process.  

The momentum scale $\mu$ in Eq.~(\ref{lagrangianEFT}), somewhat artificially, divides the short-distance physics encoded in the Wilson coefficients and the long-distance physics that corresponds to the computation of matrix elements of effective operators $Q_i^{(n)}(\mu)$. The process of relating higher-energy operators to lower-energy operators, which involves removing non-propagating degrees of freedom, is called matching. Depending on the problem, the matching can be done perturbatively or non-perturbatively. 

Since the EFT method is designed to capture the effects of all heavy NP particles, the construction of an effective Lagrangian at each order in $n$ can be done by choosing a particular power counting scheme and writing the most general basis of operators consistent with the symmetries of the theory. This procedure is non-trivial, especially at higher orders in $n$, where the number of possible operators grows rapidly. Minimal bases of the operators in the Standard Model Effective Field Theory (SMEFT) that are invariant under the SM gauge group have been partially constructed for the operators of dimensions 6,7, and 8~\cite{Petrov:2016azi}.
Some higher-dimensional operators describing the baryon-number violating interactions at low energy have also been constructed \cite{Weldon:1980gi,Ozer:1982qh,Arnold:2012sd,Petrov:2016azi}, as they represent the leading order at which those interactions can be encoded. 
Anomalous dimensions of those operators, which determine perturbative relations between operators at different scales, have also been computed for the dimension 6 operators in SMEFT \cite{Jenkins:2013zja,Jenkins:2013wua,Alonso:2013hga}. While the SMEFT power counting scheme is the most straightforward, with the importance of operators encoded in the operator dimension, other power counting schemes have also been explored \cite{Buchalla:2013rka}.

If the NP particles are light, i.e., their masses are comparable to or smaller than the energy scales at which an experiment is performed, they must be explicitly included in the construction of effective operators. This is often the case for the Dark Sector models containing axions, dark photons, right-handed neutrinos, etc. Explicit operators of the lowest dimension coupling such particles to the SM fields have been classified and go under the name of portals \cite{rpf6}.

Relating an effective Lagrangian to a physical process usually involves computations of process-dependent matrix elements of effective operators encoded in that Lagrangian. This often represents the most delicate part of the calculation. 

Some matrix elements can be perturbatively computed, such as those involving only weakly-interacting particles. Examples include processes with leptons in the initial and final states,  atomic transitions, or muonium-antimuonium oscillations, where the leading-order matrix elements can be computed in QED. Non-perturbative QED effects are reasonably under control, with the relevant wave functions obtained from the analytic or numerical solutions of the Schrodinger equation. It must be pointed out, however, that further improvements are required in the treatment of QED corrections to many processes, including those involving strongly-interacting particles \cite{https://doi.org/10.48550/arxiv.2208.05403}. 

The most challenging processes involve strongly-interacting particles, where the computation of the relevant matrix elements yields the largest theoretical uncertainties. Theoretical methods employed to deal with such matrix elements largely depend on the physical system under consideration. They include lattice QCD, various QCD sum-rule techniques, phenomenological fits, and quark model techniques \cite{petrov2022}. The importance of systematically-improvable computational techniques is recognized in the theory community. In the previous decade, this led to extensive development of lattice QCD techniques, which represent numerical solutions of QCD in Euclidean space. Lattice QCD calculations achieved remarkable precision for some of the quantities representing matrix elements of quark currents. For example, decay constants and semileptonic form-factors for pions and kaons are known with sub-percent accuracy \cite{USQCD:2022mmc}; similarly precise results are available for the decay constants of $B_{(s)}$ mesons \cite{https://doi.org/10.48550/arxiv.2208.05403,USQCD:2022mmc}. In addition, the lattice-QCD community is studying increasingly more complicated quantities, such as nonlocal matrix elements, multi-hadron matrix elements, and inclusive decay rates \cite{USQCD:2022mmc,Davoudi:2022bnl}.
At the percent-level many other phenomenologically relevant results have been obtained, including quark masses, bag parameters for neutral meson mixing, and form factors for semileptonic decays of heavy mesons and baryons, with further improvements expected in the near future \cite{USQCD:2022mmc,Boyle:2022uba}.

While the model-independence of other methods, such as QCD sum rule techniques, is often debated, they represent an approximate solution of QCD in Euclidean space and are based on the operator product expansion techniques \cite{Khodjamirian:2020btr}, which require more independent inputs at increased precision levels. The issues often raised concern the stability of sum rules with respect to the so-called Borel parameter or reliance on various versions of quark-hadron duality. Nevertheless, QCD sum rule analyses often provided estimates of physics observables long before lattice results were available \cite{Khodjamirian:2020btr}.

\begin{figure}[ht]
    \includegraphics[height=4.5in]{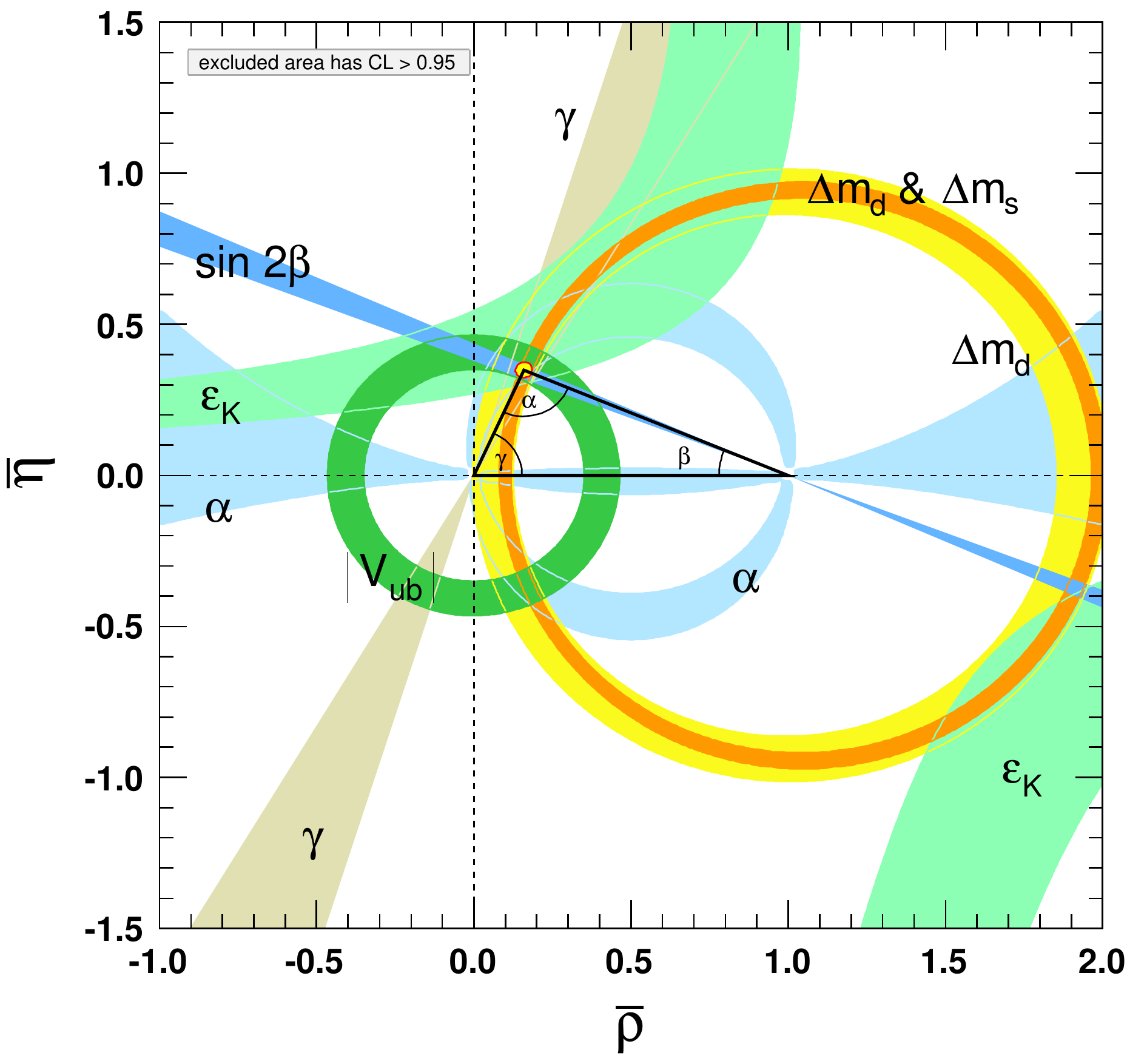}
    \caption{Various indirect studies of New Physics can be combined to provide constraints on both SM and BSM parameters. An example is studies of the CKM unitarity triangle \cite{Workman:2022ynf}
    , which requires precise knowledge of nonperturbative parameters.}
    \label{fig:theory}
\end{figure}

Sometimes the internal symmetries of the system allow for additional simplifications. For example, transitions involving heavy quark systems, mesons, and baryons involve several momentum scales, such as the scale associated with the mass of the heavy quark, $m_Q$, and the QCD intrinsic scale $\Lambda_{\rm QCD}$. The methods of heavy quark effective theory (HQET) allow us to relate various matrix elements in the heavy quark limit by expanding them in the ratio $\Lambda_{\rm QCD}/m_Q$~\cite{https://doi.org/10.48550/arxiv.2208.05403,https://doi.org/10.48550/arxiv.2207.14594}. Higher order corrections associated with such expansion often depend on soft matrix elements \ref{fig:theory} that can be computed in lattice QCD, QCD sum rules, or fit to experimental data \cite{https://doi.org/10.48550/arxiv.2208.05403}. 

Treatments of more complicated multi-scale problems with heavy quarks have also been developed and applied to quarkonium systems (Non-Relativistic QCD (NRQCD)) and non-leptonic decays of heavy hadrons (Soft-Collinear Effective Theory (SCET)). Other internal symmetries, such as isospin or flavor SU(3), can also be employed to reduce the number of unknown parameters. One should point out that the resummation methods of SCET have also been applied to study QED effects, in particular to soft or collinear photons that can lead to large logarithms and to hadronic structure-dependent effects in leptonic and semileptonic transitions \cite{https://doi.org/10.48550/arxiv.2208.05403}. 

Computations of transitions involving the light-quark systems can sometimes be simplified by applying chiral symmetry arguments \cite{Goudzovski:2022scl}. Chiral Perturbation Theory ($\chi$PT) techniques allow for controllable expansions of transition matrix elements in powers of $E/\Lambda_\chi$, where $E$ is the energy scale associated with the transition, and $\Lambda_\chi \simeq 1$ GeV is the chiral symmetry breaking scale in QCD. External parameters of $\chi$PT, the low energy constants, can be obtained from the experimental fits or direct lattice QCD calculations \cite{Goudzovski:2022scl}.   

While some of the experimental observables, such as anomalous magnetic moments $a_\ell$, leptonic electric dipole moments $d_\ell$, or lepton-flavor violating transitions among leptons, do not involve strongly-interacting particles in the initial or final state,  their effects do come into computations of perturbative electroweak corrections. For example, a theoretically-controlled calculation of the hadronic vacuum polarization (HVP) contribution to the anomalous magnetic moment of the muon $a_\mu$, which constitutes the largest theoretical uncertainty in computing this quantity, is crucial in understanding the source of the disparity between the theoretical predictions and experimental measurements \cite{https://doi.org/10.48550/arxiv.2209.08041}. The results of sophisticated computations of HVP using continuum methods \cite{Aoyama:2020ynm} are currently in tension with the direct lattice QCD calculation by the BMW collaboration \cite{Borsanyi:2020mff}. It is expected that new, precise results for HVP from other lattice collaborations as well as new experimental measurements of the low-energy $e^+e^-$ hadronic cross sections will shed light on this tension in the coming years \cite{Colangelo:2022jxc}. 

The largest theoretical uncertainties affect the interpretation of experimental data involving nuclei. Sometimes, symmetry arguments can be used to remove the computations of nuclear matrix elements, as in the case of transitions between $0^+$ nuclei used to extract the CKM matrix element $V_{ud}$. However, explicit computations are still required for the sub-leading effects \cite{Goudzovski:2022scl}. In general, the computations of nuclear effects are done with the help of nuclear models. Unfortunately,  such computations can only provide an estimate of theoretical uncertainties. Examples of such processes include muon-to-electron conversion $\mu^-  N \to e^-  N$~\cite{https://doi.org/10.48550/arxiv.2209.00142} or neutrinoless double beta decay $A (Z) \to A(Z+2) e^-e^-$ \cite{https://doi.org/10.48550/arxiv.2208.00010}. In both examples, even the observation of the transition would indicate the presence of New Physics.

Theoretical research also plays an important role in the computations of background processes. Two such calculations that will be important are the evaluations of radiative pion and radiative muon capture in the Mu2e and COMET experiments. Both processes are irrelevant for NP searches but constitute backgrounds that need to be understood. The radiative pion capture process can have the same experimental signature as muon-to-electron conversion $\mu^- N \to e^- N$. Radiative muon capture has an endpoint well below the muon-to-electron conversion signal but is a potential problem for measurements of the $\Delta L = 2$ process $\mu^-N \rightarrow e^+ N^{\prime}(Z-2)$.

The community recognizes the value of multi-institution theory efforts connecting high-energy physics and nuclear physics communities; the Neutrino Theory Network is one such program that is accelerating progress~\cite{ntn}. Such multi-institutional  efforts are  common  in collaborations working with lattice formulations of QCD on high-performance computers. Some examples of such collaborations include the Fermilab Lattice, HPQCD, MILC, and RBC-UKQCD Collaborations. The majority of US physicists working in lattice formulations of QCD are also members of the USQCD collaboration, whose software resources are open-source and widely used by the worldwide community \cite{https://doi.org/10.48550/arxiv.2208.05403}. Different lattice groups also cooperate to provide important inputs for experimental measurements, e.g., the Flavour Lattice Averaging Group (FLAG)~\cite{Aoki:2021kgd}.  

Another important synergy is based on collaborative interactions between the theory community and experimental collaborations. Such activities have proven to be important engines for both communities. There are several examples; in particular we note the Heavy Flavor Averaging Group (HFLAV), that periodically provides updates of properties of heavy-flavored states and their transitions \cite{HFLAV:2022pwe}, the Muon g-2 Theory Initiative \cite{Aoyama:2020ynm}, and the Joint Physics Analysis Center (JPAC) at Indiana University studying properties of exotic states of the hadrons \cite{JPAC}.   

Continued support for leadership in dark-sector theory is also critical both for developing models to address open problems in particle physics and cosmology, and for maximizing the efficacy of the dark-sector experimental program \cite{rpf6}.

Indirect studies of BSM physics currently provide the most stringent constraints on the possible scale of New Physics, as well as several observables whose values exhibit intriguing deviations from the SM expectations. A robust US theory program, which supports both researchers and computing resources, is essential for properly interpreting such experiments. 

\section{Connections with Other Fields\label{sec:connections}}

All the Frontiers have connections to other fields 
inside and outside particle physics.
Mechanical engineering,
electrical engineering, and materials science 
are integral components of the experimental program in particle physics, where state-of-the-art  technologies continue to be tools of discovery.
Advances in computer science are critical data processing and data mining in experimental particle physics and their interpretation with precise theoretical interpretation and prediction tools. The diversity of 
topics and techniques in the Rare and Precision 
Frontier leads to a multifaceted web of profound connections. Here we focus on a few examples to illustrate how more
cross-disciplinary efforts would benefit 
HEP.  

\subsection{Connection With Other Frontiers}
\begin{description}
\item[Energy Frontier] First, the Energy Frontier hopes to produce particles signaling new physics that might explain observations in other Frontiers.  For example, the discovery of SUSY or leptoquarks could explain a discovery in charged lepton flavor violation, the $B$-anomalies, or the muon $g {-}2$ result. Many such examples could be added.  The breadth of capabilities of the general-purpose Energy Frontier detectors allows them to pursue some measurements that are synergistic with RPF studies.   For example, DM particles can be produced in high-energy collisions in energy ranges complementary to RPF experiments focusing on masses between  1 MeV/c$^2$ -- 1 GeV/$c^2$~\cite{doi:10.1146/annurev-nucl-101917-021008}. New experiments such as CODEX-b, Faser 2, or MATHUSLA have been proposed to use high-energy collisions to search for long-lived dark matter 
particles~\cite{https://doi.org/10.48550/arxiv.2209.07156}. Rare $b$ hadron decays with final states involving charged muons can be measured with high precision. Specific hadron spectroscopy studies can be carried out at either $e^+e^-$ colliders or Energy Frontier $pp$ machines~\cite{https://doi.org/10.48550/arxiv.2207.14594}.

\item[Neutrino Physics Frontier]
There are two areas of connection to the Neutrino Frontier.  First, finding $0\nu2\beta$ would tell us that neutrinos are Majorana particles. Complex neutrino Majorana phases can yield leptogenesis, which can then produce baryogenesis~\cite{doi:10.1146/annurev.nucl.55.090704.151558}. Although we cannot predict the matter-antimatter asymmetry from Majorana phases, finding that neutrinos are Majorana particles combined with a non-zero CP $\delta$ would be suggestive. Current neutrinoless double-beta decay experiments  provide a lower limit of $3 \times 10^{25}$ years~\cite{Adams_2020} and are covered in our topical group of Baryon and Lepton Number Violation. Second, there is a natural overlap between neutrino physics and charged lepton flavor violation, another of our topical groups.  Here, the simplicity of the Standard Model does not explain obvious questions: why do the neutral leptons (neutrinos) change flavor when their charged partners apparently do not?  If there is mixing, what is the mixing matrix that corresponds  to the PMNS matrix for the charged leptons, and what is its structure?  In a broader context, masses and mixing parameters among quark and leptons constitute a  set of SM parameters with patterns and hierarchies that we do not understand. It is a hope that the concerted efforts of the RPF and NF frontiers will lead to a unified picture of the underlying physics. Finally, the dark sector efforts described in this report can leverage detector techniques relevant to the neutrino frontier. In addition, dark bosons may couple to neutrinos, making it possible for neutrino physics experiments to elucidate this sector about which so much is not yet known.


\item[Cosmic Frontier] The most immediate connection to the Cosmic Frontier comes from the search for dark matter, which is a focus of the RPF.  In addition, the search for lepton and baryon number violation in heavy baryon oscillations  or new sources of CP violation can provide  clues on the origin of the baryon asymmetry of the universe. As we will see in the discussion of connections to AMO and nuclear physics, RPF connects to AMO through interferometry that can search for gravitational waves, certainly a topic of the Cosmic Frontier.  

\item[Theory Frontier] Theory is essential in every experimental frontier to provide an intellectual structure, motivations for new experiments, calculations, and interpretation of experimental results.  We wish to stress two connections. Effective Field Theories  are especially useful in RPF experiments, since the RPF focuses on indirect searches for physics at unknown mass scales. EFTs are a natural method for categorizing contributions from different sources and interpreting data~\cite{petrov2022,Boughezal:2022cbl}. The other special connection for our Frontier is Lattice QCD.  Lattice QCD calculations relate our studies of all the hadron decays to fundamental quantities such as absolute values of CKM angles and predict SM expectations for rare decays, evaluate matrix elements relevant to the $ g {-} 2$ anomaly, and predict mass spectra of hadron multiplets \cite{Davoudi:2022bnl}. They are essential to extract fundamental properties with systematically improvable and well-understood theoretical uncertainties.

\item[Accelerator Frontier] RPF especially connects to the Accelerator Frontier subgroup ``Accelerators for Rare Processes and Physics Beyond Colliders."~\cite{https://doi.org/10.48550/arxiv.2209.06289}  Restricting our comments to that subgroup, a new muon program using PIP-II requires a new and unique fixed-field alternating gradient synchrotron (FFA), leading to a future muon program that explores low-energy muon physics and charged lepton flavor violation.

\item[Instrumentation Frontier] Advances in instrumentation are essential for progress in several of our topics. 

The instrumentation frontier is organized into ten topical groups:
\begin{itemize}
    \item Quantum sensors, which leverage quantum phenomena to make measurements by manipulating quantum states, entanglement, superposition, and similar approaches. 
    \item Photon detectors, which covers photon detection at all wavelengths, from gamma ray, visible, and ultimately radio.
    \item Solid State Detectors and Tracking covers solid state sensors for tracking detectors, from silicon to diamond and other alternative materials. It also covers non-solid state trackers such as straw trackers for high-intensity experiments.
    \item Trigger and data acquisition systems, which are tasked with collecting data from the front-end devices, reducing the data volume through selection algorithms, and producing high-level quantities for further processing.
    Their scope includes  technologies to transform electrical to optical signals and synchronization mechanisms for the precise time alignment of different components.
    \item Micropattern and gas detectors, which recent developments and future needs for Micro-Pattern Gaseous Detector (MPGD) technologies, driven by the availability of modern photolithographic techniques. 
    \item Calorimetry, namely the measurement of particle energy, in particular the energy of photons and $\pi ^0$'s.
    \item Electronics and ASICs for signal processing and data aggregation. The goal is to allow higher active element density, enhanced performance, lower power consumption, lower mass, higher radiation tolerance, or performance at cryogenic temperature.
    \item Noble elements, covering  the use of noble elements for particle detection in the gas, liquid, or solid phases, as well as other similar techniques. Included technologies include TPCs at large and small scales, which  include the measurement of scintillation, ionization, and other interaction signatures. 
    \item Cross-cutting and system integration, examining the connections between the instrumentation components summarized here and different experimental efforts, as well as considering the aspect involved with large system integration and the infrastructure needed to carry out this work, such as test beam facilities or irradiation facilities.
    \item Radio-wave detection.
\end{itemize}

High-precision timing (${\cal O}$(10ps)) is an increasingly essential feature that connects tracking detectors, calorimetry, and data acquisition systems and influences the performance of several experiments considered here.
For example, the LHCb Phase II upgrade, with its high occupancy environment, leverages  the association of tracks and showers with the right primary vertex to process the events more efficiently, with a highly reduced combinatorial background. 

An overall optimization of the material budget is key to the performance of most of the systems described in the experimental section. It is epitomized by the ultimate goal of a ``transparent tracker" in future upgrades of the Mu2e experiment, sensitive to 105 MeV/c electrons or the $\mu \rightarrow e \gamma$ and $\mu \rightarrow 3e$ experiments especially sensitive to electrons near the Michel peak at 52.8 MeV/c.  These low-energy electrons suffer large multiple scattering and $dE/dx$ relative to the high-energy particles observed in many other experiments.

Particle identification is an essential feature of heavy flavor experiments. Future needs include radiation-hard, fast, and cost-effective photosensors. Some of these needs are shared by dark sector experiments.

This highly abbreviated description is only meant to showcase the profound connection between the physics goals of the RPF frontier and the goals of the instrumentation frontier to develop new detector elements, front-end electronics, and trigger and data acquisition systems. Many of these technologies affect the performance of detectors planned for many different frontiers. New investments and long-term commitment to advances in instrumentation, along with the exploitation of synergies and the development of interdisciplinary research efforts, are essential for progress in all of particle physics.

\item[Computational Frontier] Computing connects to all the Frontiers, and RPF has many connections to computing:
\begin{itemize}
    \item Lattice QCD is an essential tool across our Frontier.  Computing time is a significant issue in improving Lattice calculations. An important question a Lattice theorist often will ask about a prediction is ``how much computing time did it take?" as part of their evaluation of the quality of the calculation.  Current simulation goals require more than an order-of-magnitude improvement in performance.\cite{https://doi.org/10.48550/arxiv.2204.00039}
    \item There are many detector and physics simulation challenges that need to be addressed, summarized in Ref.~\cite{10.3389/fphy.2022.913510}, although this article is oriented toward Energy Frontier and heavy flavor experiments rather than low-energy experiments.  We touch on a few:
    \begin{itemize}
    \item The RPF wants to send a  strong and emphatic message,  also discussed in the Computing Frontier report: GEANT4 is not sufficiently supported in the U.S.   The physics models of some crucial processes, including but not limited to their cross-sections, rates, and spectra, are in disrepair~\cite{https://doi.org/10.48550/arxiv.2203.07645,https://doi.org/10.48550/arxiv.2203.07614}.  The core group at SLAC  has largely retired and there is no identified funding source.  Many experiments in RPF rely on low-energy phenomena whose simulations  are not kept up to date; when bugs and errors are found, they are not fixed because there is no one to fix them.  GEANT is infrastructure akin to ``roads and bridges";  the current trajectory endangers progress across particle physics. 
    \item  Heterogeneous architectures present significant challenges for HEP software.  GPU and FPGA manufacturers tend to use their own software stacks, requiring specialized knowledge.  HEP programmers are usually trained in C++. While cross-platform tools are being developed, taking advantage of the options requires re-writing large fractions of the HEP software stack.  Furthermore, most HEP code bases are not easily vectorizable or parallelizable.  Problems such as particle transport are difficult to run on GPUs because of thread divergence.  Taking advantage of the massive data flows from experiments such as LHCb (restricting ourselves to the RPF)  requires fundamental change and new resources.
    \item LHCb has moved to a largely triggerless model where offline processing is performed in real-time.  Such a system is more efficient since offline flows are pushed online, decreasing storage requirements, and allows a data taking approach that can be optimized for specific physics goals.    This model of single-pass, real-time execution is very attractive and enhances its discovery potential. It must feature robust real-time reconstruction and calibration algorithms.
    \end{itemize}
\end{itemize} 

We note a common issue that affects many of the above points: there has been a major drop in funding for permanent positions in HEP that allow researchers to work on both experiment-specific tools and general issues.   Computing Frontier Group 2 addresses these problems  for Lattice QCD, event generators, and GEANT,  and we strongly endorse their calls for action~\cite{compF}.

\item[Underground Facilities Frontier] RPF does not require new underground facilities beyond those already planned in conjunction with the Neutrino Frontier for low cosmic-ray background experiments such as neutrinoless double-beta decay searches.

\item[Community Engagement Frontier] 

The Community Engagement Frontier and RPF share the goal of improving the communication of our science to the public. High-profile analyses that can be linked to broad themes that capture the public's imagination have been very successful in connecting the general public with some flagship results. We give a few examples here. Anomalies hinting at possible new physics manifestations received a lot of attention. 
  The $g {-}2$ experiment was a rare breakthrough for our field --- the story of the ring's travels from BNL to Fermilab unquestionably engaged the public.  Discoveries of exotic hadrons such as pentaquarks and tetraquarks received much attention. In times when attention to scientific achievements may be less prominent, we need to leverage  these successes and broaden the connection between captivating themes and more specialized and abstract studies that are perhaps more difficult to explain.   A more systematic approach to developing shared material and creative new communication tools is needed to improve the connection between scientists and the public, as well as strengthen advocacy efforts to receive the necessary financial support to implement the ambitious program that we are planning.

The education and mentoring of a diverse workforce in an inclusive climate allowing students and junior scientists to develop their talents in the most nurturing environment is essential for our community to thrive. There are a few advantages in the RPF experimental environment.  The smaller collaborations and the diverse nature of our experimental portfolio create opportunities for people with different aspirations and work styles.  Collaborators largely know each other personally.  We believe it is possible for those in leadership positions to help set the culture of their experiments, and in a smaller group members can reinforce norms and apply social pressure to each other more effectively.  However, a small collaboration can also easily fall prey to unacceptable behavior if  senior members tolerate it without  institutional guardrails.  A shared goal is to develop a deeper understanding of how to use these advantages while developing common strategies to improve the climate across the various experimental communities. An important tool is the practice of periodic climate surveys. For example, the LHCb collaboration, after performing such a survey, identified critical needs of its early career scientists, and implemented several changes, one of which was the creation of the ECGD (early career and gender diversity) group~\cite{Smith:2021rtv}: the leaders of this group are available for the confidential discussion of concerns from early career scientists or related to gender diversity, and promote awareness of the collaboration of issues that affect this community.

The Frontier needs to address some issues that affect career development for Early Career researchers. For example, researchers in fields such as muon physics tend to be older and leaders tend to stay in positions rather than cycle through as in large experiments such as ATLAS or CMS. It would be facile to simply assert that term limits should be enforced -- the collaborations often do not have enough people willing or capable of taking on leadership roles for such rotations to occur regularly.    The muon $g{-}2 $ experiment is again an exception; we should  study how thisexperiment renewed itself with a younger generation of leaders. The time scale for modern experiments is certainly relevant.  It would be useful to examine other ``success stories" as well and  discern patterns of success and failure.

Lastly, the innovations needed to construct the ambitious instruments that we envision call for multiple connections with industries, especially small industries eager to launch  new ambitious projects such as the ones supported by the SBIR grants. Collaborative efforts among scientists at universities and laboratories and  industrial partners are keys to innovation that will allow us to both achieve the demanding specifications discussed in the R\&D section and examine the applicability of specific technologies to non-HEP applications.

\end{description}

\subsection{Connections with Other Fields}

\begin{description}

\item[AMO and Nuclear Physics]
~~~

The discovery of an electric dipole moment for a fundamental particle such as the electron, or more complicated systems from the proton or neutron to  atoms and molecules, would be a clear signal of new physics.  Such searches probe mass scales of $10^6$ TeV/c$^2$ or higher, far higher than those directly accessible at any conceivable collider. A coordinated, complementary set of EDM searches in AMO, nuclear, and particle physics experiments could:
(a) discriminate among mechanisms that predict the observed matter-antimatter asymmetry
and (b) determine whether CP is spontaneously or explicitly broken~\cite{https://doi.org/10.48550/arxiv.2203.08103}.

Fermion EDMs originate fromnew physics at a high mass scale, requiring new CP-violating phases.  Those elementary particle EDMs  manifest themselves in bound states at lower energy scales.For protons and neutrons, Lattice QCD is an essential tool in linking the very high-energy origins of EDMs to their manifestation in particles we can observe in the laboratory.  Chiral perturbation theory, and nuclear and atomic calculations, continue to even lower-energy scales.  Hence theory leads in one direction, but it cannot predict which of the many possibilities occurs in nature; experiment leads upwards from measurements on electrons in the laboratory to BSM physics at inaccessibly high energies. 

These low-energy experiments are often studying systems such as nucleons, atoms, and molecules are all fertile ground.  Any of these low-energy bound states produce electric fields three-to-four orders of magnitude beyond what can be produced by conventional laboratory methods, making them competitive with the neutron EDM for sensitivity to quark EDMs and $\theta_{\rm QCD}$. The intellectual overlaps are beyond question.

Improvements in AMO techniques have started to rely on creating quantum superpositions analogous to those of Quantum Information Science, certainly a growing topic in particle physics.  Advances in QIS might then improve the AMO measurements.  Finally, fundamental symmetry tests rely on new quantum sensors: interferometry, optomechanical sensors, and clocks are all cutting-edge tools.  Interferometry, in particular, could be used to make an atomic interferometer for gravitational wave detection. Dark matter could also be detected using dual-species interferometers operated with different isotopes.

Another example of overlap, this time with nuclear physics, is the physics of muon capture in muon-to-electron conversion.  The muon beams for these experiments are typically $\sim 40$ MeV/c with only a few MeV of kinetic energy to stop them in thin targets.  The muon cascades to a muonic 1s state and can then be converted to an electron in a coherent process with the nucleus.  The nuclear wave function is required to calculate the overlap of the two wave functions and then a rate.  In addition, the form factor allows differentiation among CLFV models, especially at higher $Z$ materials.  The main background driving the beam's time structure is radiative pion capture (RPC): $\pi^- N \rightarrow \gamma N^{\prime}$ where the photon can convert and produce an electron at the signal energy. Unfortunately, the RPC process has not been measured on the likely targets for any proposed muon-to-electron conversion process, and the calculations are decades out-of-date.  The RPC rate can be measured {\it in situ} in Mu2e and Mu2e-II.  

Particle production as a result of  muon capture  ($\approx 60$\%) is also not well-known.  However, entire nuclear physics experiments (such as AlCap~\cite{ AlCap:2021msk}) have been performed to determine the rates of ejected particles.  Their rates and spectra play a significant role in the design of muon-to-electron conversion experiments, since highly-ionizing particles such as protons, ejected during muon capture, can deaden detector elements.

A second mode with $\Delta L = 2$, $\mu^- N(A,Z) \rightarrow e^+N(A, Z-2)$ is related to neutrinoless double-beta decay through the Schechter-Valle  ``black-box” theorem~\cite{PhysRevD.25.2951}. Here, the radiative muon capture process $\mu^- N \rightarrow \gamma N^{\prime} \nu_{\mu}$ is the background, where the photon again converts.  The spectrum is poorly known, and nuclear calculations still use the ``closure” approximation, which cuts off the spectrum at a kinematic endpoint that may not be appropriate~\cite{ CHRISTILLIN1980331}. Furthermore, the process can proceed by either a ground-state to ground-state transition or by a transition to an excited state.  There are no modern calculations of the ratio of the ground-to-excited state transitions, and the model of the excited state is a simple giant dipole resonance.  While {\it statistically} the measurements of this mode can be improved by as many orders of magnitude as the improvement in muon-to-electron conversion, this strictly nuclear physics process will be a limitation of the measurement.  Many of the same techniques required for $0\nu2\beta$ calculations are applicable here; collaborations between nuclear physicists and high-energy physicists have had only limited success because of the division of the two disciplines.  A collaboration between the two fields will be required to make progress.

Finally, experimental and theoretical studies of the properties of the spectrum of hadrons are intimately tied to the underlying theory of QCD, and is therefore also explored in nuclear physics \cite{https://doi.org/10.48550/arxiv.2207.14594,https://inspirehep.net/literature/1398831}.

\end{description}

\section{Conclusions}
The Rare Processes and Precision Measurements Frontier, as expressed in its name, seeks new physics through searches for rare processes and  precise measurements. We have identified several common goals that span the Frontier and specific recommendations to achieve our objectives.

\begin{itemize}
\item The physics of flavor and generations is a common theme --- so much so that we believe it ought to have its own driver. We study flavor through decays of $b$, $c$, and the light quarks (RPF1 and RPF2),  measuring precisely the elements of the CKM matrix and performing unitarity tests \cite{https://doi.org/10.48550/arxiv.2208.05403,https://doi.org/10.48550/arxiv.2209.07156}. Charged lepton flavor violation (CLFV in RPF5), whether in muons, taus, or heavy states (including the Higgs), is another significant focus of the Frontier.  All of these groups study rare decays, whether in the $b$ system, CP-violation in the kaon sector, or searches for CLFV. Here we have four specific recommendations:
\begin{itemize}

\item Support the LHCb Phase 2 upgrades and the Belle II program. These experiments perform incisive tests of the Standard Model. Anomalies in the $b$ sector test the unitarity of the CKM matrix along with lepton universality in $B\rightarrow K 
\ell^+\ell^-$ processes. These experiments are not U.S.-based, but we have made significant contributions.  We should to continue our role in studying the physics and see the returns on a generation of investment.

\item Embark on a program in muon-based charged lepton flavor violation using the PIP-II accelerator at Fermilab. The intense muon beams made available will surpass any other programs in the world by orders of magnitude and cannot be duplicated. Mu2e was endorsed by the last P5; Mu2e-II is the first step in a staged program that will lead to a systematic study of muon CLFV in all modes, across a wide range of possible sources, at mass scales approaching $10^5$ TeV.

\item  Strengthen and formalize the R\&D in instrumentation that is necessary to carry out this program. Fast timing detectors, precision calorimetry, and low mass trackers are all essential. In addition, the U.S.\ has an astounding infrastructure in its national laboratories that could drive transformative  advances in instrumentation. 

\item Pursue strategic R\&D in accelerator techniques. The suggested AMF muon program requires the construction of new small rings designed around those experiments~\cite{CGroup:2022tli}.  These rings can also enable a world-class dark matter beam-dump experiment. The program also requires high power targeting in a superconducting solenoid, as high as 1 MW; this presents similar problems to those for muon collider targeting. This program could help solve this challenging  problem through R\&D at intermediate stages. A staged CLFV program from 100 kW moving upwards could help provide a reliable path to a 1 MW-class muon collider target.
\end{itemize}

\item 
Fundamental symmetry tests are central to our Frontier. The topical group RPF2, examining the physics of light quarks,  identified opportunities in $K$ and $\eta/\eta^\prime$ rare decays. CP-violation in the kaon system has become an off-shore program, centered at CERN and J-PARC~\cite{Goudzovski:2022scl,Goudzovski:2022vbt}; the U.S.\ could, if it chose, revive on-shore programs, and the U.S.\ should strengthen our participation in those efforts. Further insights into fundamental symmetries can be gained through the studies of rare pion decays, for which the interesting PIONEER experimental program has been proposed \cite{PIONEER:2022yag}, which also explores lepton universality outside of the $b$-system. In addition, $\eta,\eta^\prime$ factories, such as the upcoming JEF experiment\cite{pro-JEF14} or the proposed REDTOP experiment\cite{REDTOP:2022slw}, explore fundamental physics accessible in rare $\eta,\eta^\prime$ decays. 
\item
Discovering sources of CP-violation outside the CKM matrix would have profound implications for our studies of leptogenesis and baryogenesis. EDM measurements (RPF3) are ideal for these studies.  A proposal for a proton storage ring EDM experiment could reach $10^{-29}e\cdot$cm would result in a three -order of magnitude enhancement for the QCD $\theta$-term EDM over the current limit set by neutron EDM experiments~\cite{https://doi.org/10.48550/arxiv.2203.08103}. Discovering sources of CP-violation outside the CKM matrix would have profound implications for our studies of leptogenesis and baryogenesis. EDM measurements (RPF3) are ideal for these studies.  A proposal for a proton storage ring EDM experiment could reach $10^{-29}e\cdot$cm would result in a three order of magnitude enhancement for the QCD $\theta$-term EDM over the current limit set by neutron EDM experiments~\cite{https://doi.org/10.48550/arxiv.2203.08103}. 

\item The particle nature of dark matter remains one of the greatest mysteries in all of physics.  RPF6 has identified a wide range of experiments sensitive to different ``portals" between Standard Model and dark matter. The U.S.\ should identify a set of such experiments and pursue them aggressively.  Such experiments could be performed  in  hadronic beam dumps, in electron or muon beams, or be added to DUNE.  This exciting and rapidly developing area of research is being actively pursued at CERN; the question for the U.S.\ is whether it will be a leader or cede studies to CERN, as it has in $b$-physics.

\item Baryon and Lepton Number Violation probe fundamental symmetries in particle physics.  Our Frontier works in tandem with the large neutrino experiments to search for proton decay or neutrinoless double beta decay; we add efforts in $n\bar{n}$ oscillation and rare $\Delta L = 2$ muon processes.

\item Hadron spectroscopy, the driving force of high-energy physics in its early decades, has experienced a renaissance over the past 20 years due to the discovery of scores of new, potentially ``exotic'' states (tetraquarks, pentaquarks, hybrid mesons, glueballs), as well as the observation of many new ``conventional’’ hadrons. Such research is and will be carried out at LHCb and other LHC experiments, Belle II, BESIII, and GlueX, as well as the approved Electron-Ion Collider, the proposed Super Tau-Charm factory, and other future experiments.  

\item A robust theory program, supporting both researchers and computing resources, is essential for the success of the experimental activities described in this report.
\end{itemize}

The Rare and Precision Frontier uses the techniques of many accelerators and experiments to search for new physics and do precision SM tests with diverse and complementary approaches.  We have summarized the major themes that emerged during the Snowmass process; the range and power of RPF program are unique and several experiments that we examined encompass the flexibility that enables them to elucidate unexpected observations regardless of where they are found.  

\rhead{~}

\bibliographystyle{JHEP}
\bibliography{main}
\end{document}